\documentclass[journal]{IEEEtran}
\usepackage{cite}
\usepackage{amsmath,amssymb,amsfonts}
\usepackage{algorithm}  
\usepackage{algpseudocode}  
\usepackage{algpseudocode}
\usepackage{graphicx}
\usepackage{epstopdf}
\usepackage{textcomp}
\usepackage{xcolor}
\usepackage{bm}
\usepackage{setspace}

\ifCLASSINFOpdf
\else
\fi
\begin{document}
%
\title{Joint Communication and Computation Design in Transmissive RMS Transceiver Enabled Multi-Tier Computing Networks}

\author{Zhendong~Li,~Wen~Chen,~\IEEEmembership{Senior~Member,~IEEE},~Ziwei~Liu,~Hongying~Tang,~and~Jianmin~Lu,~\IEEEmembership{Member,~IEEE}
    \thanks{This work is supported by National key project 2020YFB1807700 and 2018YFB1801102, by Shanghai Kewei 20JC1416502 and 22JC1404000, Pudong PKX2021-D02 and NSFC 62071296.}
	\thanks{Z. Li, W. Chen, and Z. Liu are with the Department of Electronic Engineering, Shanghai Jiao Tong University, Shanghai 200240, China (e-mail: lizhendong@sjtu.edu.cn; wenchen@sjtu.edu.cn; ziweiliu@sjtu.edu.cn).}
	\thanks{H. Tang is with the Science and Technology on Microsystem Laboratory, Shanghai Institute of Microsystem and Information Technology, Chinese Academy of Sciences, Shanghai 200050, China (e-mail: tanghy@mail.sim.ac.cn).}
	\thanks{J. Lu is with the Wireless Technology Laboratory, Huawei Technologies, Shanghai 201206, China (e-mail: lujianmin@huawei.com).}
	\thanks{(\emph{Corresponding author: Wen Chen.})}}
\maketitle

\begin{abstract}
	 In this paper, a novel transmissive reconfigurable meta-surface (RMS) transceiver enabled multi-tier computing network architecture is proposed for improving computing capability, decreasing computing delay and reducing base station (BS) deployment cost, in which transmissive RMS equipped with a feed antenna can be regarded as a new type of multi-antenna system.  We formulate a total energy consumption minimization problem by a joint optimization of subcarrier allocation, task input bits, time slot allocation, transmit power allocation and RMS transmissive coefficient while taking into account the constraints of communication resources and computing resources. This formulated problem is a non-convex optimization problem due to the high coupling of optimization variables, which is NP-hard to obtain its optimal solution. To address the above challenging problems, block coordinate descent (BCD) technique is employed to decouple the optimization variables to solve the problem. Specifically, the joint optimization problem of subcarrier allocation, task input bits, time slot allocation, transmit power allocation and RMS transmissive coefficient is divided into three subproblems to solve by applying BCD. Then, the decoupled three subproblems are optimized alternately by using successive convex approximation (SCA) and difference-convex (DC) programming until the convergence is achieved. Numerical results verify that our proposed algorithm is superior in reducing total energy consumption compared to other benchmarks.
\end{abstract}


%
\IEEEpeerreviewmaketitle

\section{Introduction}
\IEEEPARstart{T}{he} continuous evolution of wireless communications has spawned many emerging applications and services, e.g., telemedicine, industrial Internet and smart Internet-of-Things (IoT) \cite{7498684,7397856}. In these computing and communication-oriented application scenarios, a large number of devices and sensors need to carry out continuous communication and computing, which greatly increases the requirements for devices and sensors. Currently, such computing and communication networks face two main challenges. First, due to the small size of these devices and sensors, the communication, storage and computing capabilities are usually limited, so they cannot support computing-intensive tasks well, which will result in large computing delays and affect users' quality-of-service (QoS) \cite{8016573}. Therefore, the key issues that the next generation communication network needs to solve are considering how to reduce the computing delay and improve the computing capability of the network. In addition, since the next-generation communication network may use higher frequency bands, the propagation loss will become larger, so its coverage will be reduced. In order to achieve the same coverage as existing communication networks, the number of base stations (BSs) deployed needs to be increased. Moreover, in order to improve the QoS of users, the BS of the next-generation communication network will adopt more antennas, which will increase the required radio frequency (RF) chains, thereby increasing the cost of a single BS. Therefore, how to reduce the deployment cost of the BS is another key challenge that needs to be solved urgently in the next-generation communication network.

{\subsection{Related Works}}
\subsubsection{MEC systems}To address the first challenge, powerful computing nodes (CNs) or mobile edge computing (MEC) servers can be deployed at the network edge (i.e. usually co-located with an access point (AP) or BS), which is the recently proposed MEC technology \cite{8063331,8274943,9599533,8630994,9676649}. This technology mainly provides cloud-like computing by deploying MEC servers distributedly in the network, and is widely regarded as an effective means to liberate mobile devices from heavy computing tasks. In the MEC system, devices and sensors with limited computing capability can offload computation-intensive and latency-sensitive tasks to nearby BSs and APs equipped with MEC servers for remote execution, which can greatly reduce computing latency \cite{8488502}. It is worth noting that the prerequisite for achieving such a goal is that the computing tasks of the devices and sensors can be successfully offloaded. However, since some devices and sensors may be located at the cell edge, their offloading rate is limited, which makes the computing delay at CN or MEC longer than local computing. As a result, these devices and sensors often have to rely on their own resources for computing, which often cannot well support applications for computation-intensive and latency-sensitive tasks. Therefore, it is imperative to improve the offloading capability from the communication perspective, thereby improving the performance of computing and communication networks.

\subsubsection{Design of MEC systems}In recent years, the design of MEC communication systems has been widely discussed in academia \cite{6574874,8693989,7442079,7542156,8387798,8496832,8463562}. Note that in the MEC system, offloading strategies play a crucial role. At present, MEC offloading strategies are mainly divided into binary offloading strategies and partial offloading strategies \cite{7879258}. The binary offloading strategy mainly decides whether computing tasks are executed locally on devices and sensors or offloaded to CNs or MEC servers for remote execution. The typical tasks used for this offloading strategy are usually simple and indivisible. The partial offloading strategy usually needs to divide the computing task into several sub-tasks, and these sub-tasks can be executed locally through devices and sensors and offloaded to the CNs or MEC servers for parallel execution. Such parallel computing can greatly improve the computing capability and reduce the computing delay of the MEC system. The typical tasks used for this offloading strategy are usually multiple fine-grained processes. In addition, since the rate of offloading will also affect the performance of the MEC system, based on the above two offloading strategies, many research works have studied the joint communication and computing resource allocation in different scenarios to improve the performance of the MEC system. 

In previous work, the research on MEC systems can be divided into single-user MEC systems \cite{6574874,8693989,7442079,7542156} and multi-user MEC systems \cite{8063331,8387798,8496832,8463562}. In single-user MEC communication systems, Zhang \emph{et al.} provided a theoretical framework for energy-optimal MEC under stochastic wireless channels by optimizing the execution of mobile applications in mobile devices (i.e., mobile execution) or offloading to the cloud (i.e., cloud execution) to save energy for mobile devices \cite{6574874}. You \emph{et al.} proposed an energy-efficient computing framework that includes a set of policies to control CPU cycles for local computing modes, energy transfer, and offload time division for other offloading modes \cite{7442079}. As for the multi-user MEC system, Ren \emph{et al.} studied the delay minimization problem in the multi-user time division multiple access system with joint communication and computing resource allocation \cite{8387798}. Dai \emph{et al.} proposed a novel two-layer computation offloading framework in heterogeneous networks. Then, in a multi-task MEC system, the joint computation offloading and user association problem is formulated to minimize the overall energy consumption \cite{8496832}. In addition to the research on the basic MEC system, some emerging communication systems assisted by the MEC are also investigated. Bai \emph{et al.} studied the application of MEC in unmanned aerial vehicle (UAV) communication networks, and designed an energy-efficient physical layer security optimization algorithm \cite{8693989}. Liu \emph{et al.} studied the application of MEC in the Internet-of-Vehicles, and introduced a vehicle fog edge computing paradigm. It is then formulated as a multi-stage Stackelberg game to be solved. However, for multi-user MEC systems, the distribution locations of devices and sensors are usually random and different. Devices and sensors located at the cell edge have a large path loss to the APs or BSs, which will consume more communication resources for offloading, resulting in uneven resource allocation and user fairness issues.

\subsubsection{RMS communication systems}Moreover, faced with the challenge of high cost of BS deployment in next-generation communication networks, reconfigurable meta-surface (RMS) may be a potential solution. RMS has recently been proposed as an emerging technology combining metamaterials and communications, which can be used to reduce network costs, improve network coverage, spectrum- and energy- efficiency \cite{8811733,9531372,9362274,9509394,9716123}. RMS consists of numerous low-cost passive units, and the amplitude and phase shift of the incident signal can be changed by artificially adjusting these units. It is worth noting that since RMS is a passive device, it only adjusts the amplitude and phase of the incident signal, so it will not introduce additional noise, which makes it well applied to a collaborative communication network \cite{9531372}. In addition, compared with the existing multi-antenna technologies equipped with a large number of RF chains, the hardware cost and power consumption required by passive RMS are much lower, which also greatly stimulates research on RMS-based multi-antenna communication systems. Overall, these advantages mentioned above have greatly promoted the application of RMS in next-generation communication networks \cite{9117136,8930608}.

\subsubsection{Design of RMS enabled communication systems}Based on the above advantages, RMS has attracted extensive attention in academia and industry. Nowadays, RMS is mainly used to assist and enable traditional communication networks, where by combining active-passive beamforming design, the performance of the network can be improved with reflective or transmissive RMS \cite{9474428,9464248,9326394,9203956,9365009,9200683,9491943,9690478,9629335}. First, some works on reflective RMS-assisted communication networks has been carried out. Ur Rehman \emph{et al.} addressed the joint active and passive beamforming optimization problem under ideal and practical IRS phase shifts for an IRS-assisted multi-user downlink MIMO system, where the spectrum efficiency is maximized by minimizing the sum mean squared error (MSE) of the user's received symbols \cite{9474428}. Zen \emph{et al.} considered an IRS-assisted uplink non-orthogonal multiple access (NOMA) system in which a semi-definite relaxation technique is employed to maximize the sum rate of users \cite{9203956}. Furthermore, the research on transmissive RMS-assisted communication networks is also in progress. Zhang \emph{et al.} proposed an intelligent omni-surface (IOS) assisted downlink communication system, where the IOS is able to forward the received signal to the user in a reflection or transmission manner, thereby enhancing the wireless coverage \cite{9200683}. Niu \emph{et al.} investigated a MIMO network assisted by reflection-transmission reconfigurable intelligent surface (RIS), where the weighted sum rate is maximized based on an energy splitting (ES) scheme \cite{9629335}. Furthermore, in addition to assisting and enabling traditional communication networks, the RMS can also act as transceivers in communication networks. Since the reconfigurability of the RMS helps it to expand the number of passive units without increasing the number of expensive and bulky active antennas, the reflective RMS equipped with an active feed antenna can be used as a new type of transmitter \cite{9133266}. Since the feed blockage of the transmissive RMS transceiver is less than that of the reflective RMS transceiver, the aperture efficiency can be designed to be higher, and the operating bandwidth can be designed to be more stable, so it is more efficient \cite{bai2020high,9570775}. At present, some work on the uplink and downlink transmissive RMS transceiver design has been carried out \cite{9570775,li2021uplink}, but it is still in its infancy. Meanwhile, the application of the transmissive RMS transceiver in various communication scenarios also has potential.

\subsection{Motivation and Contributions}
Based on the above backgrounds and challenges, in order to improve the computing capability, reduce the computing delay, and reduce the BS deployment cost of the communication and computing network, we propose a transmissive RMS transceiver enabled multi-tier computing networks, where the decoding-and-forward (DF) relay is equipped with a CN, and transmissive RMS transceiver is equipped with an MEC server. In general, the computing capability of the DF relay should be comparable to or greater than that of the device to make computational cooperation feasible. To the best of our knowledge, the current research on communication and computing networks with transmissive RMS transceivers is still in its infancy, and the demand for improving network computing capability, reducing computing delay, and reducing BS deployment cost has greatly promoted this work. In this paper, we minimize total energy consumption by jointly optimizing the subcarrier allocation, task input bits, time slot allocation, transmit power allocation, and RMS transmissive coefficient. It is challenging to address this non-convex optimization problem due to the high coupling of optimization variables. Hence, we need to design an effective optimization algorithm for solving it.
In summary, the main contributions of this paper can be summarized as follows:
\begin{itemize}
	\item We propose a novel transmissive RMS transceiver enabled multi-tier computing framework, where the devices and sensors can offload tasks to DF relay and RMS multi-antenna system for calculations, thereby improving computing capability and reducing computing latency of the networks. Meanwhile, we formulate the energy consumption minimization problem for joint optimization of the subcarrier allocation, task input bits, time slot allocation, transmit power allocation, and RMS transmissive coefficient. Since the objective function and the partial constraints are non-convex due to the high coupling of the optimization variables, the problem is a non-convex optimization problem and is challenging to solve directly.
	
	\item To address the formulated energy consumption minimization problem, we first divide the non-convex optimization problem into three sub-problems based on the block coordinate descent (BCD) algorithm. Specifically, in the first sub-problem, given the time allocation, task input bits, and RMS transmissive coefficient, we solve the joint optimization problem for the subcarrier allocation and user transmit power allocation. In the second sub-problem, we deal with the joint optimization problem for the time allocation and task input bits by applying successive convex approximation (SCA) when the subcarrier allocation, user transmit power allocation and RMS transmissive coefficient are fixed. For the third sub-problem, the RMS transmissive coefficient can be obtained by using difference-convex (DC) programming and SCA when other optimization variables are given. Finally, the three sub-problems are optimized alternately until convergence is achieved.
	
	\item Through the numerical simulation, we verify the effectiveness of the proposed joint optimization algorithm for the subcarrier allocation, task input bits, time slot allocation, transmit power allocation and RMS transmissive coefficient compared with the benchmark algorithms, i.e., it can decrease the total energy consumption. In addition, the proposed multi-layer offload-computation scheme is superior to other offload schemes, and the introduction of transmissive RMS transceivers can greatly reduce the cost of overall network deployment, which has great potential in next-generation communications.
\end{itemize}

The rest of this paper is organized as follows. Section II elaborates the system model and optimization problem formulation for the transmissive RMS transceiver enabled multi-tier computing networks. Section III presents the proposed optimization algorithm for the formulated optimization problem. In Section IV, numerical results demonstrate that our algorithm has good convergence and effectiveness. Finally, conclusions are given in Section V.

\emph{Notations:} Scalars are denoted by lower-case letters, while vectors and matrices are represented by bold lower-case letters and bold upper-case letters, respectively. $\left| {x} \right|$ denotes the absolute value of a complex-valued scalar $x$. For a square matrix $\bf{X}$, $\rm{tr(\bf{X})}$, $\rm{rank(\bf{X})}$, ${\bf{X}}^H$ ,${\bf{X}}_{m,n}$ and $\left\| {\bf{X}} \right\|$ denote its trace, rank, conjugate transpose, $m,n$-th entry and matrix norm, respectively, while ${\bf{X}} \succeq 0$ represents that $\bf{X}$ is a positive semidefinite matrix. Similarly, for a general matrix $\bf{A}$, $\rm{rank(\bf{A})}$, ${\bf{A}}^H$, ${\bf{A}}_{m,n}$ and $\left\| {\bf{A}} \right\|$ also denote its rank, conjugate transpose, $m,n$-th entry and matrix norm, respectively. In addition, ${\mathbb{C}^{M \times N}}$ denotes the space of ${M \times N}$ complex matrices. ${\bf{I}}_N$ denotes an dentity matrix of size ${N \times N}$. $j$ denotes the imaginary unit, i.e., $j^2=-1$. $\mathbb{E}\left\{  \cdot  \right\}$ represents the expectation of random variables. Finally, the distribution of a circularly symmetric complex Gaussian (CSCG) random vector with mean $\mu$ and covariance matrix $\bf{C}$ is denoted by $ {\cal C}{\cal N}\left( {\mu,\bf{C}} \right)$, and $\sim$ stands for `distributed as'.
\section{System Model and Problem Formulation}
In this section, we mainly describe the system model and problem formulation.
\subsection{Network Model}
As shown in the Fig. 1, we consider a multi-tier MEC network model based on a relay-transmissive RMS multi-antenna system, which includes $K$ single-antenna task nodes (TN), a single-antenna DF relay and $M$ transmissive elements RMS multi-antenna system. In this paper, we consider the orthogonal frequency division multiple access (OFDMA) system, where the channel of bandwidth $B$ is divided into $N$ subcarriers, each with a bandwidth of $W = {B \mathord{\left/
		{\vphantom {B N}} \right.
		\kern-\nulldelimiterspace} N}$. Inter-subcarrier interference is negligible, and the cyclic prefix is large enough to overcome inter-symbol interference. Note that TN $k$ needs to successfully execute the $D_k>0$ task input bits in the time duration $T>0$. We consider that TN $k$, DF relay, and RMS multi-antenna systems are all computationally capable. Specifically, the $D_k$ task input bits of TN $k$ can be divided into three parts for computation: local computation, offloading to DF relay computation, and offloading to RMS multi-antenna system computation. Let $d_k^l$, $d_k^r$,  $d_k^m$ denote the number of task input bits for TN $k$ to be computed locally, offloaded to the DF relay computation, and offloaded to the RMS multi-antenna system computation, respectively. Thus, we have
\begin{equation}
d_k^l + d_k^r + d_k^m = {D_k},\forall k.
\end{equation}
\begin{figure}
	\centerline{\includegraphics[width=8cm]{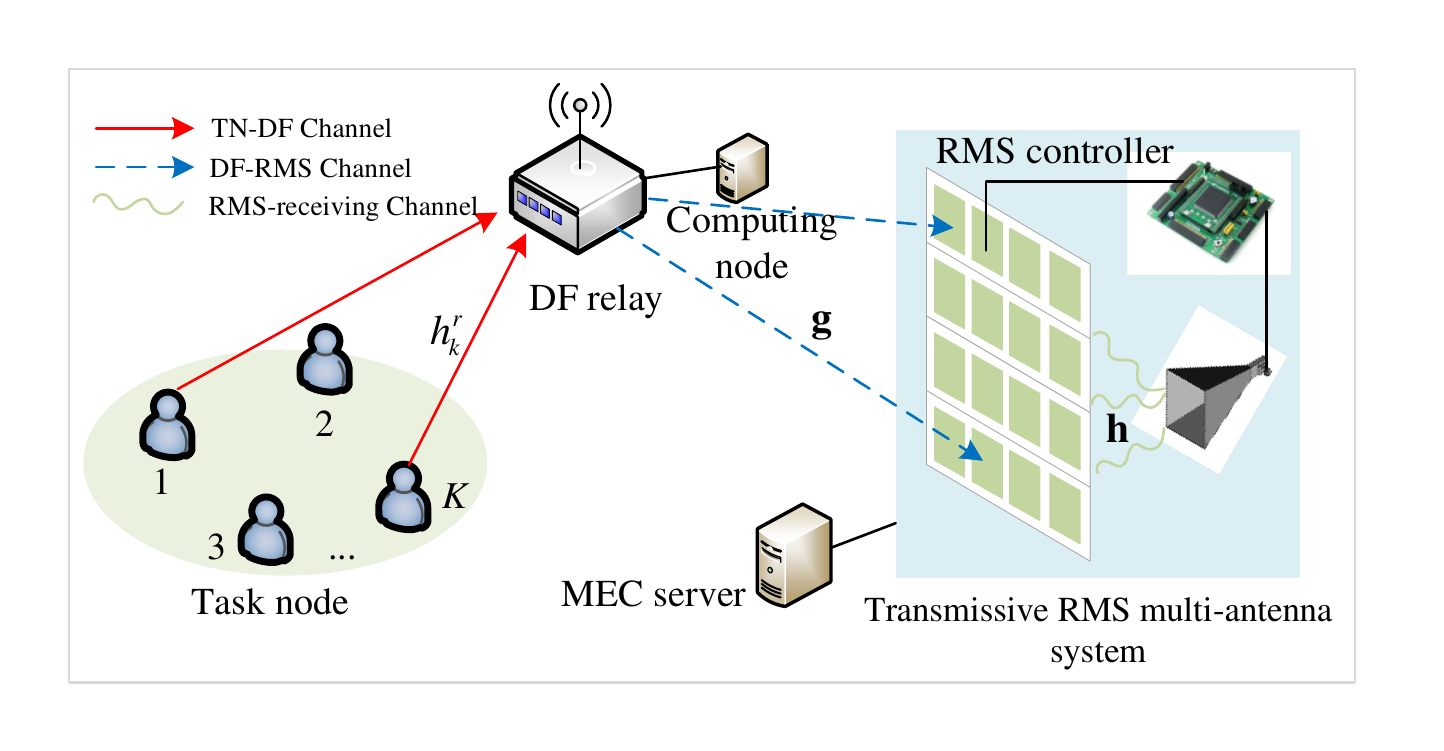}}
	\caption{Transmissive RMS transceiver enabled multi-tier computing networks.}
	\label{Fig1}
\end{figure}
\subsection{Multi-tier Offloading-Computing Model}
We divide the time duration $T$ of TN $k$ for offloading and computing into four-time slots as shown in Fig. 2. In the first time slot $t_k^{\rm{{\rm I}}} \ge 0$, TN $k$ offloads $d_k^r$ task input bits to the DF relay. We assume that the CN and the DF relay are co-located and connected by using high-throughput low-latency optical fibers, so the data transmission between the DF relay and CN can be considered delay-free. The CN executes the task for the remaining $T - t_k^{\rm I}$ time. In the $t_k^{{\rm I}{\rm I}} \ge 0$ and $t_k^{{\rm I}{\rm I}{\rm I}} \ge 0$ time slots, TN $k$ offloads the task input bits $d_k^m$ to the RMS multi-antenna system through the DF relay. Specifically, in the second time slot $t_k^{{\rm I}{\rm I}}$, TN $k$ sends the task input bits $d_k^m$ to the DF relay. After successfully decoding the received $d_k^m$, the DF relay forwards it to the RMS multi-antenna system within the third time slot $t_k^{{\rm I}{\rm I}{\rm I}}$. The RMS multi-antenna system receives and decodes the signal and sends it to the MEC server. The MEC server executes these tasks in the fourth time slot $t_k^{{\rm I}{\rm{V}}} \ge 0$. We still assume that the MEC and RMS multi-antenna systems are co-located and connected using high-throughput low-latency fiber, so data transmission between the two can also be considered delay-free.

Since the calculation result bits is usually much smaller than the calculation input bits, the time for the user to download the calculation result is negligible compared to the time for offloading. To ensure that the tasks of TN $k$ can be successfully executed within time duration $T$, we have the following constraints:
\begin{equation}
t_k^{\rm I} + t_k^{{\rm I}{\rm I}} + t_k^{{\rm I}{\rm I}{\rm I}} + t_k^{{\rm{{\rm I}V}}} \le T,\forall k.
\end{equation}
\begin{figure}
	\centerline{\includegraphics[width=8cm]{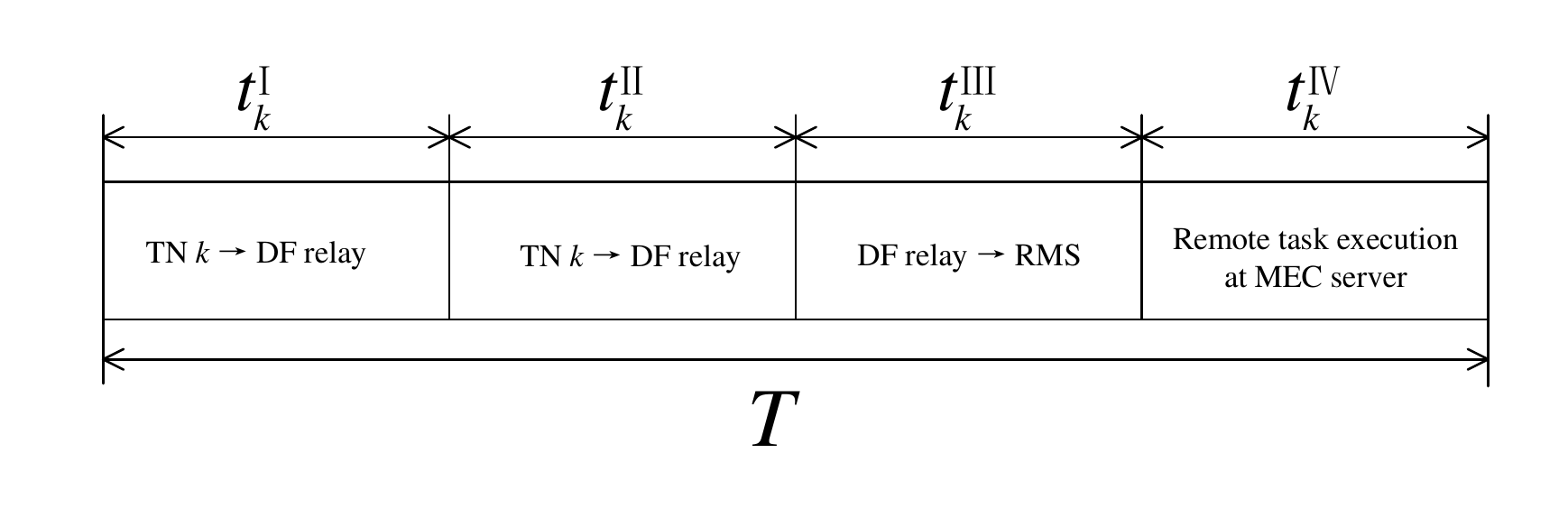}}
	\caption{Multi-tier offloading-computing model.}
	\label{Fig1}
\end{figure}
\subsection{Offloading Model}
\subsubsection{Offload to DF relay}
In the time slot $t_k^{\rm I}$, TN $k$ offloads $d_k^r$ task input bits to the DF relay with transmit power $P_{k,n}^{\rm I}$. Then the achievable data rate for task offloading of TN $k$ to DF relay on $n$-th subcarrier can be expressed as
\begin{equation}
r_{k,n}^{\rm I} = {a_{k,n}}W{\log _2}\left( {1 + \frac{{P_{k,n}^{\rm I}{{\left| {h_{k,n}^r} \right|}^2}}}{{{\sigma ^2}}}} \right),\forall k,n,
\end{equation}
where ${\sigma ^2}$ represents the power of additive white Gaussian noise (AWGN) introduced at the DF relay. ${a_{k,n}} \in \left\{ {0,1} \right\}$ indicates whether the $n$-th subcarrier is allocated to TN $k$. When the $n$-th subcarrier is allocated to TN $k$, ${a_{k,n}} = 1$. Otherwise, ${a_{k,n}} = 0$. It should satisfy the following constraints
\begin{equation}
\sum\limits_{k = 1}^K {{a_{k,n}}}  \le 1,\forall n,
\end{equation}
and $h_{k,n}^r$ represents the channel gain from the TN $k$ to DF relay on $n$-th subcarrier, which can be modeled as
\begin{equation}
h_{k,n}^r = \sqrt {\frac{{{C_0}}}{{d_k^\nu }}} \left( {\sqrt {\frac{{{\kappa _1}}}{{1 + {\kappa _1}}}} {e^{ - j2\pi nW\frac{{{d_k}}}{c}}} + \sqrt {\frac{1}{{1 + {\kappa _1}}}} {{\tilde h}_{k,n}}} \right),\forall k,n,
\end{equation}
with ${\tilde h_{k,n}} \sim {\cal C}{\cal N}\left( {0,1} \right)$, where $C_0$ represents the channel gain when the reference distance is 1m, $d_k$ represents the distance from the TN $k$ to the DF relay, $\nu$ denots the path loss coefficient of the corresponding channel, and ${\kappa _1}$ represents the Rician factor of the channel corresponding to TN $k$.  Therefore, in time slot $t_k^{\rm I}$, the task input bits $d_k^r$ of TN $k$ offloaded to the DF relay can be expressed as
\begin{equation}
d_k^r = t_k^{\rm I}\sum\limits_{n = 1}^N {r_{k,n}^{\rm I}} ,\forall k.
\end{equation}
Let $P_{\max }^t$ denote the maximum transmit power of TN $k$, so we have
\begin{equation}
P_{k,n}^{\rm I} \ge 0,\forall k,n,
\end{equation}
and
\begin{equation}
\sum\limits_{n = 1}^N {{a_{k,n}}P_{k,n}^{\rm I}}  \le P_{\max }^t,\forall k.
\end{equation}

For this offloading process, we regard the transmission energy consumption of TN as the main energy consumption and ignore the energy consumption of circuits such as its radio frequency chain and baseband signal processing. Therefore, in time slot $t_k^{\rm I}$, the energy consumption of TN $k$ offloading $d_{k,n}^r$ task input bits to the DF relay on the $n$-th subcarrier can be expressed as
\begin{equation}
E_{k,n}^{\rm I} = {a_{k,n}}P_{k,n}^{\rm I}t_k^{\rm I},\forall k,n.
\end{equation}

\subsubsection{Offlaod to RMS multi-antenna system}
In the second slot $t_k^{{\rm I}{\rm I}}$ and the third slot $t_k^{{\rm I}{\rm I}{\rm I}}$, the DF relay offloads the task input bits $d_k^m$ of TN $k$ to the RMS multi-antenna system. Let $P_{k,n}^{{\rm I}{\rm I}}$ denote the transmit power of TN $k$ on the $n$-th subcarrier in the time slot $t_k^{{\rm I}{\rm I}}$, which should satisfy
\begin{equation}
P_{k,n}^{{\rm I}{\rm I}} \ge 0,\forall k,n,
\end{equation}
and
\begin{equation}
\sum\limits_{n = 1}^N {{a_{k,n}}P_{k,n}^{{\rm I}{\rm I}}}  \le P_{\max }^t,\forall k.
\end{equation}
Therefore, the achievable data rate from the TN $k$ to DF relay on the $n$-th subcarrier at time slot $t_k^{{\rm I}{\rm I}}$ can be expressed as
\begin{equation}
r_{k,n}^{{\rm I}{\rm I}} = {a_{k,n}}W{\log _2}\left( {1 + \frac{{P_{k,n}^{{\rm I}{\rm I}}{{\left| {h_{k,n}^r} \right|}^2}}}{{{\sigma ^2}}}} \right),\forall k,n.
\end{equation}

According to the characteristics of the DF relay, after successfully decoding the received signal, it forwards the signal to the RMS multi-antenna system in the third time slot $t_k^{{\rm I}{\rm I}{\rm I}}$. Specifically, the RMS multi-antenna system is composed of the receiving antenna and the transmissive RMS. On the $n$-th subcarrier, the channel from the RMS to the receiving antenna can be expressed as ${{\bf{h}}_n} \in {\mathbb{C}^{M \times 1}}$, and the channel from the DF relay to the RMS can be given by ${{\bf{g}}_n} \in {\mathbb{C}^{M \times 1}}$. We name them the receiving-RMS channel and the RMS-DF channel, respectively. In this paper, we assume that the channel is constant during the coherence time duration $T$. Therefore, we only need to obtain three sets of CSI before each optimization, and then apply these CSI in the time duration $T$. Since the DF relay has the ability to receive and transmit, some classical channel estimation algorithms can be well applied to obtain the CSI from TN $k$ to the DF relay. Then, the DF relay can transmit the obtained CSI to the controller of the RMS transceiver, and the controller performs centralized control. For DF relay to RMS and RMS to receiving antenna, since the receiving antenna has the receiving ability, the transmissive RMS has no receiving ability, the CSI of the cascaded channel can be obtained by drawing on some reflective RMS channel estimation schemes \cite{9475488,8879620}. For the channel from the DF relay to the RMS multi-antenna system, the transmissive RMS adjusts the phase and amplitude of the DF relay forwarded signal and sends it to the receiving antenna. Let ${\bf{s}} = {\left[ {{s_1},...,{s_M}} \right]^T} \in {\mathbb{C}^{M \times 1}}$ denote the transmissive coefficient vector, where ${s_m} = {\beta _m}{e^{j{\theta _m}}},\forall m$. ${\beta _m} \in \left[ {0,1} \right]$ and ${\theta _m} \in \left[ {0,2\pi } \right)$ represent the transmissive amplitude and phase shift of the $m$-th element, respectively. The transmissive coefficient ${s_m}$ needs to satisfy 
\begin{equation}
\left| {{s_m}} \right| \le 1,\forall m.
\end{equation}
For the channel model, the RMS-DF channel is the far-field channel, and the receiving-RMS channel is the near-field channel \cite{li2021uplink}. We consider the transmissive elements of the RMS to be arranged in uniform planar array (UPA), i.e. $M = {M_c} \times {M_r}$, ${M_c}$ and ${M_r}$ denote the number of RMS elements on the column and row, respectively. For the RMS-DF channel, we consider that it has both line-of-sight (LoS) and non-line-of-sight (NLoS) components, so we model it as a Rician fading channel, which can be expressed as
\begin{equation}
{\bf{g}}_n\! = \! \sqrt {\frac{{{C_0}}}{{{d^\alpha }}}} \! \left(\!  {\sqrt {\frac{\kappa _2}{{1 \! + \! \kappa _2}}} {e^{ - j2\pi nW\frac{d}{c}}}{{\bf{g}}_{{\rm{LoS}}}}\!  + \! \sqrt {\frac{1}{{1\!  + \! \kappa _2}}} {{\bf{g}}_{{\rm{NLoS}}}}} \! \right),
\end{equation}
with ${{\bf{g}}_{{\rm{NLoS}}}} \sim {\cal C}{\cal N}\left( {{\bf{0}},{{\bf{I}}_{{M_c}{M_r}}}} \right)$. ${{\bf{g}}_{{\rm{LoS}}}}$ can denoted by
\begin{equation}
\begin{aligned}
{{\bf{g}}_{{\rm{LoS}}}} = {\left[ {1,{e^{ - j2\pi {f_c}\frac{{{d_r}\sin \varphi \cos \psi }}{c}}},...,{e^{ - j2\pi {f_c}\left( {{M_r} - 1} \right)\frac{{{d_r}\sin \varphi \cos \psi }}{c}}}} \right]^T}\\\otimes {\left[ {1,{e^{ - j2\pi {f_c}\frac{{{d_c}\sin \varphi \sin \psi }}{c}}},...,{e^{ - j2\pi {f_c}\left( {{M_c} - 1} \right)\frac{{{d_c}\sin \varphi \sin \psi }}{c}}}} \right]^T},
\end{aligned}
\end{equation}
where $d$ denotes the distance between the DF relay and RMS, $\alpha$ represents the path loss coefficient of the RMS-DF channel, $c$ represents the speed of light, $\kappa $ represents the Rician factor and ${f_c}$ represents the carrier frequency. $\varphi$ and $\psi$ are the vertical and horizontal angles-of-arrival (AoA) of the incident signal at transmissive RMS, respectively.

Considering that there is no occlusion between the RMS and the receiving antenna, we model the receiving-RMS channel in the near field as a LoS channel, which can be expressed as
\begin{equation}
{{\bf{h}}_n} = \rho {e^{ - j2\pi nW\frac{{\hat r}}{c}}}{{\bf{h}}_{{\rm{LoS}}}},
\end{equation}
with
\begin{equation}
{{\bf{h}}_{{\rm{LoS}}}} = {\left[ {{e^{ - j2\pi {f_c}\frac{{{r_{1,1}} - \hat r}}{c}}},...,{e^{ - j2\pi {f_c}\frac{{{r_{{M_c},{M_r}}} - \hat r}}{c}}}} \right]^H},
\end{equation}where $\rho$ and $\hat{r}$ represent the complex channel gain and the distance from the RMS center to the receiving antenna, respectively. The distance from the $\left( {{m_c},{m_r}} \right)$-th RMS element to the receiving antenna is
\begin{equation}
{r_{{m_c},{m_r}}} = \sqrt {{{\hat r}^2} + \hat d_{{m_c},{m_r}}^2} ,
\end{equation}
where ${\hat d_{{m_c},{m_r}}} = \sqrt {\delta _{{m_c}}^2d_c^2 + \delta _{{m_r}}^2d_r^2} $ denotes the distance from the $\left( {{m_c},{m_r}} \right)$-th element of the RMS to the RMS center. $d_c$ and $d_r$ denote the column spacing and row spacing from the $\left( {{m_c},{m_r}} \right)$-th element to the center element, respectively. ${\delta _{{m_c}}} = \frac{{2{m_c} - {M_c} - 1}}{2}$ and ${\delta _{{m_r}}} = \frac{{2{m_r} - {M_r} - 1}}{2}$.

Therefore, in the time slot $t_k^{{\rm I}{\rm I}{\rm I}}$, the achievable rate of the signal forwarded by the DF relay received by the receiving antenna is
\begin{equation}
r_{k,n}^{{\rm I}{\rm I}{\rm I}} = {b_{k,n}}W{\log _2}\left( {1 + \frac{{P_{k,n}^{{\rm I}{\rm I}{\rm I}}{{\left| {{\bf{h}}_n^H{\rm{diag}}\left( {{{\bf{g}}_n}} \right){\bf{s}}} \right|}^2}}}{{{\delta ^2}}}} \right),\forall k,n,
\end{equation}
where ${\delta ^2}$ represents the power of AWGN introduced at the feeding antenna, ${b_{k,n}} \in \left\{ {0,1} \right\}$ denotes the subcarrier allocation variable, which is constrained by the following
\begin{equation}
\sum\limits_{k = 1}^K {{b_{k,n}}}  \le 1,\forall n,
\end{equation}
and $P_{k,n}^{{\rm I}{\rm I}{\rm I}} $ represents the TN $k$ transmit power allocated on the $n$-th subcarrier by the DF relay in the time slot $t_k^{{\rm I}{\rm I}{\rm I}}$, which should satisfy
\begin{equation}
P_{k,n}^{{\rm I}{\rm I}{\rm I}} \ge 0,\forall k,n,
\end{equation}
and
\begin{equation}
\sum\limits_{k = 1}^K {\sum\limits_{n = 1}^N {{b_{k,n}}P_{k,n}^{{\rm I}{\rm I}{\rm I}}}  \le P_{\max }^r} ,
\end{equation}
where $P_{\max }^r$ denotes the maximum transmit power of DF relay.

According to the achievable rate of TN $k$ in the time slot $t_k^{{\rm I}{\rm I}}$ and $t_k^{{\rm I}{\rm I}{\rm I}}$, the task input bits $d_k^m$ of TN $k$ offloaded to RMS multi-antenna system through DF relay needs to satisfy
\begin{equation}
d_k^m = \min \left( {t_k^{{\rm I}{\rm I}}\sum\limits_{n = 1}^N {r_{k,n}^{{\rm I}{\rm I}}} ,t_k^{{\rm I}{\rm I}{\rm I}}\sum\limits_{n = 1}^N {r_{k,n}^{{\rm I}{\rm I}{\rm I}}} } \right),\forall k.
\end{equation}
We consider the transmission energy consumption of TN $k$ and DF relay for offloading as the main energy consumption in the time slot $t_k^{{\rm I}{\rm I}}$ and $t_k^{{\rm I}{\rm I}{\rm I}}$. We have
\begin{equation}
E_{k,n}^{{\rm I}{\rm I}} = {a_{k,n}}P_{k,n}^{{\rm I}{\rm I}}t_k^{{\rm I}{\rm I}},\forall k,n,
\end{equation}
and
\begin{equation}
E_{k,n}^{{\rm I}{\rm I}{\rm I}} = {b_{k,n}}P_{k,n}^{{\rm I}{\rm I}{\rm I}}t_k^{{\rm I}{\rm I}{\rm I}},\forall k,n.
\end{equation}
Therefore, the offloading energy consumption of TN $k$ offloaded to the DF relay and to the RMS multi-antenna system through the DF relay can be expressed as
\begin{equation}
E_k^{off} = \sum\limits_{n = 1}^N {\left( {E_{k,n}^{\rm I} + E_{k,n}^{{\rm I}{\rm I}} + E_{k,n}^{{\rm I}{\rm I}{\rm I}}} \right)} ,\forall k.
\end{equation}

\subsection{Computing Model}
\subsubsection{TN local computing model}
During the time duration $T$, TN $k$ executes $d_k^l$ task input bits. In fact, the number of CPU cycles to perform a computing task depends largely on various factors, e.g., the specific application, the number of task input bits, and the hardware device used for the computation. To characterize the most necessary computation and communication tradeoff, we consider the number of CPU cycles to execute a task as a linear function of the number of task input bits, where $c_t$ represents the number of CPU cycles to compute each task input bit at TN. Let ${f_{t,i}}$ denote the CPU frequency of the $i$-th cycle of TN, which is subject to the following constraints:
\begin{equation}
{f_{t,i}} \le {f_{t,\max }},\forall i \in \left\{ {1,...,{c_t}d_k^l} \right\},
\end{equation}
where ${f_{t,\max }}$ represents the maximum CPU frequency when the TN executes the task. Since the task input bits computed locally by TN $k$ should be successfully executed within time duration $T$, we have the following delay constraints:
\begin{equation}
\sum\limits_{i = 1}^{{c_t}d_k^l} {\frac{1}{{{f_{t,i}}}}}  \le T.
\end{equation}
Therefore, the local computing energy consumption of TN $k$ can be expressed as
\begin{equation}
E_k^{t,comp} = \sum\limits_{i = 1}^{{c_t}d_k^l} {{\alpha _t}f_{t,i}^2} ,
\end{equation}
where ${\alpha _t}$ depends on the effective capacitance factor of TN chip architecture. In order to save computing power consumption under computating latency constraints, it is better to set the CPU frequency to be the same for different CPU cycles \cite{8234686}. By using this fact and making the constraints in Eq. (28) satisfy strict equality (in order to minimize computing energy consumption), we have
\begin{equation}
{f_{t,1}} = {f_{t,2}} = ... = {f_{t,{c_t}d_k^l}} = \frac{{{c_t}d_k^l}}{T},\forall k.
\end{equation}
Therefore, the local computing energy consumption of TN $k$ can be further expressed as
\begin{equation}
E_k^{t,comp} = \frac{{{\alpha _t}{{\left( {{c_t}d_k^l} \right)}^3}}}{{{T^2}}},\forall k.
\end{equation}
Additionally, the CPU frequency for local computation is bound by the maximum CPU frequency that can be given by
\begin{equation}
\frac{{{c_t}d_k^l}}{T} \le {f_{t,\max }},\forall k.
\end{equation}

\subsubsection{CN computing model}
After receiving the $d_k^r$ task input bits offloaded by TN $k$ in the first time slot $t_k^{\rm I}$, the DF relay executes the calculation through the connected CN. We assume that the two are co-located and connected using high-throughput, low-latency fiber, so their transmission delays are negligible. CN executes tasks within the remaining $T - t_k^{\rm I}$. Let ${f_{r,i}}$ and ${f_{r,\max }}$ denote the CPU frequency and the maximum frequency of the CPU in the $i$-th cycle of CN, respectively. Similar to the local computing of TN $k$, we have
\begin{equation}
{f_{r,i}} = \frac{{{c_r}d_k^r}}{{T - t_k^{\rm I}}},\forall i \in \left\{ {1,...,{c_r}d_k^r} \right\},
\end{equation}
where $c_r$ represents the number of CPU cycles that CN executes each task input bit. Therefore, the computing energy consumption of CN to execute $d_k^r$ task bits offloaded by TN $k$ can be expressed as
\begin{equation}
E_k^{r,comp} = \frac{{{\alpha _r}{{\left( {{c_r}d_k^r} \right)}^3}}}{{{{\left( {T - t_k^{\rm I}} \right)}^2}}},\forall k,
\end{equation}
where ${\alpha _r}$ depends on the effective capacitance factor of CN chip architecture. In order to ensure that the CN can successfully execute the task, the task input bits should satisfy the following constraints
\begin{equation}
\sum\limits_{k = 1}^K {\frac{{{c_r}d_k^r}}{{T - t_k^{\rm I}}}}  \le {f_{r,\max }},
\end{equation}
where ${f_{r,\max }}$ represents the maximum CPU frequency when CN executes tasks.

\subsubsection{MEC computing model}
In the fourth time slot $t_k^{{\rm{{\rm I}V}}}$, the RMS multi-antenna system receives the $d_k^m$ task input bits offloaded of TN $k$ by the DF relay on the $n$-th subcarrier, and then forwards it to the nearby MEC server for computing. Similarly, we assume that the two are co-located and connected using high-throughput, low-latency fiber, so their transmission delays are also negligible. Therefore, the time required for the MEC server to execute $d_k^m$ task input bits is expressed as
\begin{equation}
t_k^{{\rm{{\rm I}V}}} = \frac{{{c_m}d_k^m}}{{{f_{m,i}}}},\forall i \in \left\{ {1,...,{c_r}d_k^m} \right\},
\end{equation}
where $c_m$ represents the number of CPU cycles that MEC executes each task input bit. Similarly, in order to ensure that the MEC can successfully execute the task, there are the following constraints on the task input bits
\begin{equation}
\sum\limits_{k = 1}^K {\frac{{{c_m}d_k^m}}{{t_k^{{\rm{{\rm I}V}}}}}}  \le {f_{m,\max }}.
\end{equation}
\subsection{Problem Formulation}
Considering that TN and DF relay are usually wireless devices, the energy management of them is more complicated, and the RMS transceiver as a BS usually has a reliable power supply, so we temporarily consider the energy consumption of wireless device side (i.e., TN and DF relay) as the main performance indicator. In this paper, we aim to minimize the total energy consumption of TN and DF relay in the transmissive RMS transceiver enabled multi-tier computing networks by jointly optimizing the subcarrier allocation \textbf{A (B)}, task input bits \textbf{D}, time slot allocation \textbf{T}, transmit power allocation \textbf{P} and RMS transmissive coefficient \textbf{s}. This optimization problem can be formulated as
\begin{subequations}
	\begin{align}
	({\rm{P}}0){\rm{ ~~~~}}&\mathop {\min }\limits_{{\bf{A}},{\bf{B}},{\bf{D}},{\bf{T}},{\bf{P}},{\bf{s}}} {\rm{  }}\sum\limits_{k = 1}^K {\left( {E_k^{off} + E_k^{t,comp} + E_k^{r,comp}} \right)} ,\\
	s.t.\qquad &d_k^r \le t_k^{\rm I}\sum\limits_{n = 1}^N {r_{k,n}^{\rm I}} ,\forall k,\\
	&d_k^m \le \min \left( {t_k^{{\rm I}{\rm I}}\sum\limits_{n = 1}^N {r_{k,n}^{{\rm I}{\rm I}}} ,t_k^{{\rm I}{\rm I}{\rm I}}\sum\limits_{n = 1}^N {r_{k,n}^{{\rm I}{\rm I}{\rm I}}} } \right),\forall k,\\
	&d_k^l + d_k^r + d_k^m = {D_k},\forall k,\\
	&P_{k,n}^S \ge 0,S \in \left\{ {{\rm I},{\rm I}{\rm I},{\rm I}{\rm I}{\rm I}} \right\},\forall k,n,\\
	&\sum\limits_{n = 1}^N {{a_{k,n}}P_{k,n}^S}  \le P_{\max }^t,S \in \left\{ {{\rm I},{\rm I}{\rm I}} \right\},\forall k,\\
	&\sum\limits_{n = 1}^N {\sum\limits_{k = 1}^K {{b_{k,n}}P_{k,n}^{{\rm I}{\rm I}{\rm I}}} }  \le P_{\max }^r,\\
	&{a_{k,n}},{b_{k,n}} \in \left\{ {0,1} \right\},\forall k,n,\\
	&\sum\limits_{k = 1}^K {{a_{k,n}}}  \le 1,\forall n,\\
	&\sum\limits_{k = 1}^K {{b_{k,n}}}  \le 1,\forall n,\\
	&\frac{{{c_t}d_k^l}}{T} \le {f_{t,\max }},\forall k,\\
	&\sum\limits_{k = 1}^K {\frac{{{c_r}d_k^r}}{{T - t_k^{\rm I}}}}  \le {f_{r,\max }},\\
	&\sum\limits_{k = 1}^K {\frac{{{c_m}d_k^m}}{{t_k^{{\rm{{\rm I}V}}}}}}  \le {f_{m,\max }},\\
	&t_k^S \ge 0,S \in \left\{ {{\rm I},{\rm I}{\rm I},{\rm I}{\rm I}{\rm I},{\rm{{\rm I}V}}} \right\},\forall k,\\
	&t_k^{\rm I} + t_k^{{\rm I}{\rm I}} + t_k^{{\rm I}{\rm I}{\rm I}} + t_k^{{\rm{{\rm I}V}}} \le T,\forall k,\\
	&\left| {{s_m}} \right| \le 1,\forall m,
	\end{align}
\end{subequations}
where (38b)-(38d) represent the task input bit constraints for TN $k$ offload to DF relay and RMS multi-antenna system. (38e)-(38g) denote the transmit power constraints for TN $k$ and DF relay. (38h)-(38j) represent subcarrier allocation constraints in different time slots. (38k)-(38m) denote the computing capability constraints of TN $k$, CN, and MEC server. (38n) and (38o) represent time allocation constraints. (38p) represents the transmissive coefficient constraint of the RMS multi-antenna system. Note that this joint optimization problem is designed offline, i.e., we obtain the optimization variable $\left\{{{\bf{A}},{\bf{B}},{\bf{D}},{\bf{T}},{\bf{P}},{\bf{s}}} \right\}$ under the assumption that the user location and all channel state information (CSI) are all perfectly acquired.

However, the solution to the problem (P0) is challenging for the following reasons. Firstly, the optimization variables are highly coupled, which leads to the objective function and constraints being non-convex with respect to (w.r.t) the optimization variables. Then, the constraints (38f)-(38j) involve binary variables. Therefore, the problem (P0) is a mixed-integer non-convex optimization problem, and it is quite challenging to obtain the global optimal solution. Consequently, we need to design an efficient algorithm to obtain a high-quality sub-optimal solution through the BCD algorithm. The algorithm for solving the problem (P0) will be introduced in detail below.
\section{Joint Optimization Algorithm for the Transmissive RMS Transceiver Enabled Multi-Tier Computing Networks}
In this section, since the formulated problem (P0) is a non-convex optimization problem, we divide the problem (P0) into three sub-problems for solving based on the BCD algorithm. Specifically, In the sub-problem 1, the time slot allocation \textbf{T}, the task input bit \textbf{D} and the RMS transmissive coefficient \textbf{s} are fixed, and the subcarrier allocation \textbf{A}, \textbf{B} and the transmit power allocation \textbf{P} are jointly optimized. In the sub-problem 2, given the subcarrier allocation \textbf{A}, \textbf{B}, the transmit power \textbf{P} and the RMS transmissive coefficient \textbf{s}, the task input bit \textbf{D} and time allocation \textbf{T} are jointly optimized. The third sub-problem is to optimize the RMS transmissive coefficient \textbf{s} for fixed subcarrier allocation \textbf{A}, \textbf{B}, transmit power allocation \textbf{P}, task input bit \textbf{D} and time allocation \textbf{T}. Finally, the three sub-problems are optimized alternately until convergence is achieved.

The existence of binary variables \textbf{A} and \textbf{B} makes constraints (38f)-(38j) non-convex constraints. In order to solve this problem, we first relax the binary variable ${a_{k,n}}$ to obtain ${\tilde a_{k,n}}$, i.e., ${a_{k,n}} \in \left\{ {0,1} \right\} \Rightarrow {\tilde a_{k,n}} \in \left[ {0,1} \right],\forall k,n$. Then, the auxiliary variables $\tilde P_{k,n}^{\rm I} = {\tilde a_{k,n}}P_{k,n}^{\rm I},\forall k,n$ and $\tilde P_{_{k,n}}^{{\rm{II}}} = {\tilde a_{k,n}}P_{_{k,n}}^{{\rm{II}}},\forall k,n$ are introduced. Similarly, for the binary variable ${b_{k,n}}$, we have ${b_{k,n}} \in \left\{ {0,1} \right\} \Rightarrow {\tilde b_{k,n}} \in \left[ {0,1} \right],\forall k,n$, and then introduce the auxiliary variable $\tilde P_{k,n}^{{\rm{{\rm I}II}}} = {\tilde b_{k,n}}P_{_{k,n}}^{{\rm{{\rm I}II}}},\forall k,n$. After variable relaxation and the introduction of auxiliary variables, the optimization problem (P0) can be expressed as the optimization problem (P1) as follows
\begin{subequations}
	\begin{align}
	({\rm{P}}1){\rm{ ~~~~}}&\mathop {\min }\limits_{{\bf{\tilde A}},{\bf{\tilde B}},{\bf{D}},{\bf{T}},{\bf{\tilde P}},{\bf{s}}} {\rm{  }}\sum\limits_{k = 1}^K \left( {\sum\limits_{n = 1}^N {\left( {\tilde P_{k,n}^{\rm I}t_k^{\rm I} + \tilde P_{k,n}^{{\rm I}{\rm I}}t_k^{{\rm I}{\rm I}} + \tilde P_{k,n}^{{\rm I}{\rm I}{\rm I}}t_k^{{\rm I}{\rm I}{\rm I}}} \right)} }\right.\nonumber\\
	&~~~~~~~~~~~~~~~~~~~~~ \left.{+ E_k^{t,comp} + E_k^{r,comp} }\right),\\
	s.t.\qquad &\textrm {(38b)-(38d), (38k)-(38p)},\\
	&\tilde P_{k,n}^S \ge 0,S \in \left\{ {{\rm I},{\rm I}{\rm I},{\rm I}{\rm I}{\rm I}} \right\},\forall k,n,\\
	&\sum\limits_{n = 1}^N {\tilde P_{_{k,n}}^S}  \le P_{\max }^t,S \in \left\{ {{\rm I},{\rm I}{\rm I}} \right\},\forall k,\\
	&\sum\limits_{n = 1}^N {\sum\limits_{k = 1}^K {\tilde P_{_{k,n}}^{{\rm I}{\rm I}{\rm I}}} }  \le P_{\max }^r,\\
	&{{\tilde a}_{k,n}},{{\tilde b}_{k,n}} \in \left[ {0,1} \right],\forall k,n,\\
	&\sum\limits_{k = 1}^K{{\tilde a}_{k,n}}\le 0,\forall n,\\
	&\sum\limits_{k = 1}^K{{\tilde b}_{k,n}}\le 0,\forall n.
	\end{align}
\end{subequations}

\subsection{Sub-problem 1: Joint Optimization of Subcarrier Allocation and Transmit Power Allocation}
In this subsection, firstly, given the time allocation \textbf{T}, the task input bit \textbf{D} and the RMS transmissive coefficient \textbf{s}, the subcarrier allocation $\bf{\tilde A}$, $\bf{\tilde B}$ and transmit power allocation $\bf{\tilde P}$ are jointly optimized, then the problem (P1) can be written as the problem (P2), which is expressed as
\begin{subequations}
	\begin{align}
	({\rm{P}}2){\rm{ ~~~~}}&\mathop {\min }\limits_{{\bf{\tilde A}},{\bf{\tilde B}},{\bf{\tilde P}}} {\rm{  }}\sum\limits_{k = 1}^K \left( {\sum\limits_{n = 1}^N {\left( {\tilde P_{k,n}^{\rm I}t_k^{\rm I} + \tilde P_{k,n}^{{\rm I}{\rm I}}t_k^{{\rm I}{\rm I}} + \tilde P_{k,n}^{{\rm I}{\rm I}{\rm I}}t_k^{{\rm I}{\rm I}{\rm I}}} \right)}  + }\right.\nonumber\\
	&~~~~~~~~~~~~~~~~~~~~\left.{\frac{{{\alpha _t}{{\left( {{c_t}d_k^l} \right)}^3}}}{{{T^2}}} + \frac{{{\alpha _t}{{\left( {{c_r}d_k^r} \right)}^3}}}{{{{\left( {T - t_k^{\rm I}} \right)}^2}}}} \right) ,\\
	s.t.\qquad &\textrm {(38b), (39c)-(39h)},\\
	&d_k^m \le t_k^{{\rm I}{\rm I}}\sum\limits_{n = 1}^N {r_{k,n}^{{\rm I}{\rm I}}} ,\forall k,\\
	&d_k^m \le t_k^{{\rm I}{\rm I}{\rm I}}\sum\limits_{n = 1}^N {r_{k,n}^{{\rm I}{\rm I}{\rm I}}} ,\forall k,
	\end{align}
\end{subequations}
\textbf{Lemma 1:} The function $f\left( {x,t} \right) = t{\log _2}\left( {1 + \frac{x}{t}} \right)$ is concave w.r.t $x>0$ and $t>0$.

\emph{Proof:}  The function $f\left( {x,t} \right)$ can be obtained by the perspective transformation of the function $h\left( x \right) = {\log _2}\left( {1 + x} \right)$, i.e, $f\left( {x,t} \right) = t h\left( {\frac{x}{t}} \right)$. Since the perspective function is concave-preserving and $h\left( x \right) = {\log _2}\left( {1 + x} \right)$ is concave function w.r.t $x > 0$, the function $f\left( {x,t} \right) = t{\log _2}\left( {1 + \frac{x}{t}} \right)$ is concave w.r.t $x > 0$ and $t > 0$. The proof of \textbf{Lemma 1} is completed. $\hfill\blacksquare$

According to the \textbf{Lemma 1}, ${r_{k,n}^{\rm I}}$ is concave w.r.t ${{\tilde a}_{k,n}}$ and ${\tilde P_{k,n}^{\rm I}}$, ${r_{k,n}^{{\rm I}{\rm I}}}$ is concave w.r.t ${{\tilde a}_{k,n}}$ and ${\tilde P_{k,n}^{{\rm I}{\rm I}}}$, and ${r_{k,n}^{{\rm I}{\rm I}{\rm I}}}$ is concave w.r.t ${{\tilde b}_{k,n}}$ and ${\tilde P_{k,n}^{{\rm I}{\rm I}{\rm I}}}$. Hence, (38b), (40c) and (40d) are convex constraints. In addtion, the objective function is affine w.r.t optimization variables and (39c)-(39h) are also affine constraints. Thus, it can be seen that the problem (P2) is a standard convex optimization problem, which can be solved by using CVX toolbox \cite{grant2014cvx}.

\emph{Remark: To facilitate obtaining the solution to the problem (P2), the subcarrier allocation variables ${a_{k,n}}$ and ${b_{k,n}}$ are relaxed into continuous variables. After obtaining the subcarrier allocation variable value, it needs to be restored to obtain the subcarrier allocation scheme. Considering that one user can occupy multiple subcarriers, but one subcarrier can only be allocated to one user, we obtain the maximum value of all user subcarrier allocation variables on the $n$-th subcarrier and set the ${a_{k,n}}$ or ${b_{k,n}}$ of the corresponding user to 1, and the ${a_{k,n}}$ or ${b_{k,n}}$ of the remaining users to 0.}

\subsection{Sub-problem 2: Joint Optimization of Task Input Bit and Time Allocation}
In this subsection, given subcarrier allocation \textbf{A}, \textbf{B}, transmit power allocation \textbf{P} and RMS transmissive coefficient \textbf{s}, we jointly optimize task input bit \textbf{D} and time allocation \textbf{T}, then the problem (P1) can be transformed into the problem (P3), which can be expressed as
\begin{subequations}
	\begin{align}
	({\rm{P}}3){\rm{ ~~~~}}&\mathop {\min }\limits_{{\bf{D}},{\bf{T}} }{\rm{ ~ }}\sum\limits_{k = 1}^K \left( {\sum\limits_{n = 1}^N {\left( {\tilde P_{k,n}^{\rm I}t_k^{\rm I} + \tilde P_{k,n}^{{\rm I}{\rm I}}t_k^{{\rm I}{\rm I}} + \tilde P_{k,n}^{{\rm I}{\rm I}{\rm I}}t_k^{{\rm I}{\rm I}{\rm I}}} \right)}  + }\right.\nonumber\\
	&~~~~~~~~~~~~~~~~~~~~\left.{\frac{{{\alpha _t}{{\left( {{c_t}d_k^l} \right)}^3}}}{{{T^2}}} + \frac{{{\alpha _t}{{\left( {{c_r}d_k^r} \right)}^3}}}{{{{\left( {T - t_k^{\rm I}} \right)}^2}}}} \right) ,\\
	s.t.\qquad &\textrm {(38b), (38d), (38k)-(38o), (40c), (40d)}.
	\end{align}
\end{subequations}
\textbf{Lemma 2:} $E_k^{r,comp} = \frac{{{\alpha _t}{{\left( {{c_r}d_k^r} \right)}^3}}}{{{{\left( {T - t_k^{\rm I}} \right)}^2}}}$ is a convex function w.r.t $d_k^r > 0$ and $t_k^{\rm I} > 0$.

\emph{Proof:} The Hessian matrix of $E_k^{r,comp} = \frac{{{\alpha _t}{{\left( {{c_r}d_k^r} \right)}^3}}}{{{{\left( {T - t_k^{\rm I}} \right)}^2}}}$ is
\begin{equation}
\left[ {\begin{array}{*{20}{c}}
	{\frac{{6{c_r}^3{\alpha _t}d_k^r}}{{{{\left( {T - t_k^{\rm I}} \right)}^2}}}}&{\frac{{6{c_r}^3{\alpha _t}{{\left( {d_k^r} \right)}^2}}}{{{{\left( {T - t_k^{\rm I}} \right)}^3}}}}\\
	{\frac{{6{c_r}^3{\alpha _t}{{\left( {d_k^r} \right)}^2}}}{{{{\left( {T - t_k^{\rm I}} \right)}^3}}}}&{\frac{{6{c_r}^3{\alpha _t}{{\left( {d_k^r} \right)}^3}}}{{{{\left( {T - t_k^{\rm I}} \right)}^4}}}}
	\end{array}} \right].
\end{equation}
Its eigenvalues are $0$ and $\frac{{6{c_r}^3{\alpha _t}d_k^r}}{{{{\left( {T - t_k^{\rm I}} \right)}^2}}}\left( {1 + \frac{{{{\left( {d_k^r} \right)}^2}}}{{{{\left( {T - t_k^{\rm I}} \right)}^2}}}} \right)$, so the matrix is a semi-positive definite matrix, so $E_k^{r,comp}$ is jointly convex w.r.t $d_k^r > 0$ and $t_k^{\rm I} > 0$. This completes the proof of \textbf{Lemma 2}.$\hfill\blacksquare$

In the objective function of problem (P3), it can be proved that $E_k^{off}$ is affine w.r.t optimization variables, and $E_k^{t,comp}$ is convex w.r.t optimization variables. According to \textbf{Lemma 2}, $E_k^{r,comp}$ is also a convex function w.r.t the optimization variables. Therefore, the objective function of this problem (P3) is convex. In addition, (38b), (38d), (38k), (38n), (38o), (40c) and (40d) are affine. (38l) and (38m) are non-convex constraints, which makes the problem (P3) still a non-convex optimization problem. Next, we can apply SCA to carry out the first-order Taylor expansion of the left-hand-side (LHS) concave functions of constraints (38l) and (38m), and obtain their upper bounds respectively\footnote{According to the second-order condition of convex functions, the LHS of Eq. (38l) and Eq. (38m) are concave functions, which are easy to prove and are omitted here.}. 

Specifically, for the LHS ${\frac{{{c_r}d_k^r}}{{T - t_k^{\rm I}}}}$ of constraint (38l), we adopt SCA to obtain its upper bound, which can be expressed as
\begin{equation}
\begin{aligned}
\frac{{{c_r}d_k^r}}{{T - t_k^{\rm I}}} \le \frac{{{c_r}d_k^{r\left( i \right)}}}{{T - t_k^{{\rm I}\left( i \right)}}} + \frac{{{c_r}}}{{T - t_k^{{\rm I}\left( i \right)}}}\left( {d_k^r - d_k^{r\left( i \right)}} \right) +\\ \frac{{{c_r}d_k^{r\left( i \right)}}}{{{{\left( {T - t_k^{{\rm I}\left( i \right)}} \right)}^2}}}\left( {t_k^{\rm I} - t_k^{{\rm I}\left( i \right)}} \right) \buildrel \Delta \over = {\left( {\frac{{{c_r}d_k^r}}{{T - t_k^{\rm I}}}} \right)^{ub}},
\end{aligned}
\end{equation}
where $d_k^{r\left( i \right)}$ and $t_k^{{\rm I}\left( i \right)}$ represent the values of $d_k^r$ and $t_k^{\rm I}$ at the $i$-th SCA iteration, respectively.
Similarly, for the LHS ${\frac{{{c_m}d_{k,n}^m}}{{t_k^{{\rm{{\rm I}V}}}}}}$ of constraint (38m), we also adopt SCA to obtain its upper bound, which can be expressed as
\begin{equation}
\begin{aligned}
\frac{{{c_m}d_{k,n}^m}}{{t_k^{{\rm{{\rm I}V}}}}} \le \frac{{{c_m}d_{k,n}^{m\left( i \right)}}}{{t_k^{{\rm{{\rm I}V}}\left( i \right)}}} + \frac{{{c_m}}}{{t_k^{{\rm{{\rm I}V}}\left( i \right)}}}\left( {d_{k,n}^m - d_{k,n}^{m\left( i \right)}} \right) -\\ \frac{{{c_m}d_{k,n}^{m\left( i \right)}}}{{{{\left( {t_k^{{\rm{{\rm I}V}}\left( i \right)}} \right)}^2}}}\left( {t_k^{{\rm{{\rm I}V}}} - t_k^{{\rm{{\rm I}V}}\left( i \right)}} \right) \buildrel \Delta \over = {\left( {\frac{{{c_m}d_{k,n}^m}}{{t_k^{{\rm{{\rm I}V}}}}}} \right)^{ub}},
\end{aligned}
\end{equation}
where $d_{k,n}^{m\left( i \right)}$ and $t_k^{{\rm{{\rm I}V}}\left( i \right)}$ denote the values of $d_{k,n}^m$ and $t_k^{{\rm{{\rm I}V}}}$ at the $i$-th SCA iteration, respectively. Therefore, the problem (P3) can be approximately transformed into the problem (P3.1), which can be expressed as
\begin{subequations}
	\begin{align}
	({\rm{P}}3.1){\rm{ ~~~~}}&\mathop {\min }\limits_{{\bf{D}},{\bf{T}} }{\rm{ ~ }}\sum\limits_{k = 1}^K \left( {\sum\limits_{n = 1}^N {\left( {\tilde P_{k,n}^{\rm I}t_k^{\rm I} + \tilde P_{k,n}^{{\rm I}{\rm I}}t_k^{{\rm I}{\rm I}} + \tilde P_{k,n}^{{\rm I}{\rm I}{\rm I}}t_k^{{\rm I}{\rm I}{\rm I}}} \right)}  + }\right.\nonumber\\
	&~~~~~~~~~~~~~~~~~~~~\left.{\frac{{{\alpha _t}{{\left( {{c_t}d_k^l} \right)}^3}}}{{{T^2}}} + \frac{{{\alpha _t}{{\left( {{c_r}d_k^r} \right)}^3}}}{{{{\left( {T - t_k^{\rm I}} \right)}^2}}}} \right) ,\\
	s.t.\qquad &\textrm {(38b), (38d), (38k), (38n), (38o), (40c), (40d)},\\
	&{\sum\limits_{k = 1}^K {\left( {\frac{{{c_r}d_k^r}}{{T - t_k^{\rm I}}}} \right)} ^{ub}} \le {f_{r,\max }},\\
	&{\sum\limits_{k = 1}^K {\left( {\frac{{{c_m}d_k^m}}{{t_k^{{\rm{{\rm I}V}}}}}} \right)} ^{ub}} \le {f_{m,\max }}.
	\end{align}
\end{subequations}
The problem (P3.1) is a standard convex optimization problem that can be solved by using the CVX toolbox \cite{grant2014cvx}.
\subsection{Sub-problem 3: Optimization of RMS Transmissive Coefficient}
In this subsection, we fix the subcarrier allocation \textbf{A}, \textbf{B}, the transmit power allocation \textbf{P}, the task input bit \textbf{D} and the time allocation \textbf{T}, and solve the transmissive coefficient of RMS. Since the objective function does not contain the RMS transmissive coefficient vector explicitly, the problem (P1) can be transformed into feasibility-check problem (P4), which can be expressed as
\begin{subequations}
	\begin{align}
	({\rm{P}}4){\rm{ ~~~~}}&{\rm{find~~}}{\bf{s}},\\
	s.t.\qquad &d_k^m \le t_k^{{\rm I}{\rm I}{\rm I}}\sum\limits_{n = 1}^N {r_{k,n}^{{\rm I}{\rm I}{\rm I}}} ,\forall k,\\
	&\left| {{s_m}} \right| \le 1,\forall m.
	\end{align}
\end{subequations}
It can be seen that the problem (P4) is a non-convex optimization problem. To solve this problem, we let ${\bf{v}}_n^H = {\bf{h}}_n^H{\rm{diag}}\left( {{{\bf{g}}_n}} \right) \in {\mathbb{C}^{1 \times M}}$, then ${\left| {{\bf{h}}_n^H{\rm{diag}}\left( {{{\bf{g}}_n}} \right){\bf{s}}} \right|^2} = {\left| {{\bf{v}}_n^H{\bf{s}}} \right|^2} = {\bf{v}}_n^H{\bf{s}}{{\bf{s}}^H}{{\bf{v}}_n}$. Let ${\bf{S}} = {\bf{s}}{{\bf{s}}^H} \in {\mathbb{C}^{M \times M}}$, ${\bf{S}}\succeq 0$ and ${\rm{rank}}\left( {\bf{S}} \right) = 1$. In addition, we let ${{\bf{V}}_n} = {{\bf{v}}_n}{\bf{v}}_n^H \in {\mathbb{C}^{M \times M}}$, then ${\left| {{\bf{h}}_n^H{\rm{diag}}\left( {{{\bf{g}}_n}} \right){\bf{s}}} \right|^2} = {\rm{tr}}\left( {{\bf{S}}{{\bf{V}}_n}} \right)$. Therefore, $r_{k,n}^{{\rm I}{\rm I}{\rm I}}$ can be equivalently expressed as
\begin{equation}
r_{k,n}^{{\rm I}{\rm I}{\rm I}} = {b_{k,n}}W{\log _2}\left( {1 + \frac{{P_{k,n}^{{\rm I}{\rm I}{\rm I}}{\rm{tr}}\left( {{\bf{S}}{{\bf{V}}_n}} \right)}}{{{\delta ^2}}}} \right),\forall k,n.
\end{equation}
Then the problem (P4) can be equivalently expressed as the problem (P4.1), which can be given by
\begin{subequations}
	\begin{align}
	({\rm{P}}4.1){\rm{ ~~~~}}&{\rm{find~~}}{\bf{S}},\\
	s.t.\qquad &\textrm {(46b)},\\
	&{{\bf{S}}_{m,m}} \le 1,\forall m,\\
	&{\bf{S}}\succeq 0,\\
	&{\rm{rank}}\left( {\bf{S}} \right) = 1.
	\end{align}
\end{subequations}
It can be seen that the problem (P4.1) is a non-convex optimization problem due to the existence of the non-convex rank-one constraint (46e). Next, we apply \textbf{Proposition 1} to transform constraint (46e).

\textbf{Proposition 1: }For any positive semi-definite matrix ${\bf{C}} \in {\mathbb{C}^{N \times N}}$, ${\rm{tr}}\left( {\bf{C}} \right) > 0$, the rank-one constraint can be equivalently expressed as
\begin{equation}
{\rm{rank}}\left( {\bf{C}} \right){\rm{ = 1}} \Rightarrow {\rm{tr}}\left( {\bf{C}} \right) - {\left\| {\bf{C}} \right\|_2} = 0,
\end{equation}
where ${\rm{tr}}\left( {\bf{C}} \right) = \sum\limits_{n = 1}^N {{\sigma _n}} \left( {\bf{C}} \right)$, ${\left\| {\bf{C}} \right\|_2} = {\sigma _1}\left( {\bf{C}} \right)$ represents the spectral norm of the matrix $\bf{C}$. ${\sigma _n}\left( {\bf{C}} \right)$ denotes the $n$-th largest singular value of matrix $\bf{C}$.

According to \textbf{Proposition 1}, we can transform the rank-one constraint (48e) in the problem (P4.1) into
\begin{equation}
{\rm{rank}}\left( {\bf{S}} \right){\rm{ = 1}} \Rightarrow {\rm{tr}}\left( {\bf{S}} \right) - {\left\| {\bf{S}} \right\|_2} = 0.
\end{equation}
Thus, the feasible-check problem (P4.1) can be transformed into the problem (P4.2), which can be expressed as
\begin{subequations}
	\begin{align}
	({\rm{P}}4.2){\rm{ ~~~~}}&\mathop {{\rm{min}}}\limits_{\bf{S}} {\rm{~tr}}\left( {\bf{S}} \right) - {\left\| {\bf{S}} \right\|_2},\\
	s.t.\qquad &\textrm {(46b), (48c), (48d)}.
	\end{align}
\end{subequations}
Since ${\left\| {\bf{S}} \right\|_2}$ is a convex function, the problem (P4.2) is still a non-convex optimization problem. Here, we use SCA to obtain the lower bound of ${\left\| {\bf{S}} \right\|_2}$, which can be expressed as
\begin{equation}
\begin{aligned}
{\left\| {\bf{S}} \right\|_2} \ge {\left\| {{{\bf{S}}^{\left( i \right)}}} \right\|_2} \!+\! {\rm{tr}}\left(\! {{{\bf{u}}_{\max }}\left( {{{\bf{S}}^{\left( i \right)}}} \right){{\bf{u}}_{\max }}{{\left( {{{\bf{S}}^{\left( i \right)}}} \right)}^H}\!\left( {{\bf{S}} - {{\bf{S}}^{\left( i \right)}}}\! \right)} \!\right)\\ \buildrel \Delta \over = {\left( {{{\left\| {\bf{S}} \right\|}_2}} \right)^{lb}},
\end{aligned}
\end{equation}where ${{\bf{u}}_{\max }}\left( {{{\bf{S}}^{\left( i \right)}}} \right)$ denotes the eigenvector corresponding to the largest singular value of the matrix ${{\bf{S}}^{\left( i \right)}}$, and ${{\bf{S}}^{\left( i \right)}}$ represents the value of ${\bf{S}}$ in the $i$-th SCA iteration. Therefore, the problem (P4.2) can be approximately converted to the problem (P4.3), which can be given by
\begin{subequations}
	\begin{align}
	({\rm{P}}4.3){\rm{ ~~~~}}&\mathop {{\rm{min}}}\limits_{\bf{S}} {\rm{ ~tr}}\left( {\bf{S}} \right) - {\left( {{{\left\| {\bf{S}} \right\|}_2}} \right)^{lb}},\\
	s.t.\qquad &\textrm {(46b), (48c), (48d)}.
	\end{align}
\end{subequations}
It can be seen that the problem (P4.3) is a standard SDP problem, which can be solved by using CVX toolbox \cite{grant2014cvx}.

\subsection{The Overall Joint Optimization Algorithm in Multi-tier Computing Networks}
Since multiple optimization variables are highly coupled, the original problem is a complex non-convex optimization problem. In this paper, we decouple the original problem into three sub-problems to solve through the BCD algorithm framework. Specifically, in sub-problem 1, we first relax the binary variables, and then given the task input bits \textbf{D}, the time allocation \textbf{T}, and the RMS transmissive coefficient \textbf{s}, we can obtain the subcarrier allocation $\bf{\tilde{A}}$, $\bf{\tilde{B}}$, and the transmit power allocation $\bf{\tilde{P}}$. Finally, the binary variable is restored. In sub-problem 2, given subcarrier allocation \textbf{A}, \textbf{B}, transmit power allocation \textbf{P}, and RMS transmissive coefficient \textbf{s}, we obtain task input bits \textbf{D} and time allocation \textbf{T} by applying SCA technique. In sub-problem 3, with fixed subcarrier allocation \textbf{A}, \textbf{B}, transmit power allocation \textbf{P}, task input bits \textbf{D}, and time allocation \textbf{T}, we can obtain RMS transmissive coefficient \textbf{s} by applying SCA and DC programming. Finally, the three subproblems are optimized alternately until the convergence is achieved. Based on the solution of the above subproblems, we propose a joint optimization algorithm in this multi-tier computing network, which can be summarized as\textbf{Algorithm 1}.
\begin{algorithm}[H]
	\caption{The Overall Joint Optimization Algorithm in Multi-Tier Computing Networks} 
	\begin{algorithmic}[1]
		\State Initialize ${\bf{A}}^0$, ${\bf{B}}^0$, ${\bf{P}}^0$, ${\bf{D}}^0$ , ${\bf{T}}^0$, ${\bf{s}}^0$, convergence threshold $\epsilon$ and iteration index $i = 0$.
		\Repeat
		\State Obtain subcarrier allocation ${\bf{A}}^*$, ${\bf{B}}^*$ and transmit power allocation ${\bf{P}}^*$ by solving the problem (P2).
		\State Obtain task input bits ${\bf{D}}^*$, and time allocation ${\bf{T}}^*$, by solving the problem (P3.1).
		\State Obtain RMS transmissive coefficient ${\bf{s}}^*$, by solving the problem (P4.3).
		\State Update $i=i+1$.
		\Until The fractional decrease of the objective value is below a threshold $\epsilon$.
		\State \Return The subcarrier allocation, transmit power allocation, task input bits, time allocation and RMS transmissive coefficient design scheme.
	\end{algorithmic}
\end{algorithm}

\subsection{Computational Complexity and Convergence Analysis}
\subsubsection{Computational complexity analysis}
In each iteration, the problem (P2) is solved with the computational complexity of ${\cal O}\left( {{{\left( {KN} \right)}^{3.5}}} \right)$, and the problem (P3.1) is solved with computational complexity of ${\cal O}\left( K ^{3.5}\right)$ \cite{boyd2004convex}. The problem (P4.3) solves a SDP problem by interior point method, so the computational complexity can be represented by ${\cal O}\left( {{ M ^{3.5}}} \right)$. It can be assumed that the number of iterations required for the algorithm to reach convergence is $i$, the computational complexity of the proposed algorithm can be expressed as ${\cal O}\left( {i\left( {\left(KN\right) ^{3.5}+K^{3.5} +  {{M}^{3.5}} } \right)} \right)$.

\subsubsection{Convergence analysis}
The convergence of the proposed joint subcarrier allocation, transmit power allocation, task input bits, time allocation and RMS transmissive coefficient optimization in multi-tier computing networks is elaborated as follows. 

We define ${\bf{A}}^{(i)}$, ${\bf{B}}^{(i)}$, ${\bf{P}}^{(i)}$, ${\bf{D}}^{(i)}$ , ${\bf{T}}^{(i)}$ and ${\bf{s}}^{(i)}$ as the $i$-th iteration solution of the problem (P2), (P3.1) and (P4.3). Herein, the objective function in the $i$-th iteration is denoted by ${\cal E}\left( {{{\bf{A}}^{(i)}},{{\bf{B}}^{(i)}},{{\bf{P}}^{(i)}},{{\bf{D}}^{(i)}},{{\bf{T}}^{(i)}},{{\bf{s}}^{(i)}}} \right)$. In the step 3 of $\textbf{Algorithm 1}$, since subcarrier allocation and the transmit power allocation can be obtained for given ${\bf{D}}^{(i)}$ , ${\bf{T}}^{(i)}$ and ${\bf{s}}^{(i)}$. Hence, we have 
\begin{equation}
\begin{aligned}
{\cal E}\left( {{{\bf{A}}^{(i)}},{{\bf{B}}^{(i)}},{{\bf{P}}^{(i)}},{{\bf{D}}^{(i)}},{{\bf{T}}^{(i)}},{{\bf{s}}^{(i)}}} \right) \ge \\{\cal E}\left( {{{\bf{A}}^{(i + 1)}},{{\bf{B}}^{(i + 1)}},{{\bf{P}}^{(i + 1)}},{{\bf{D}}^{(i)}},{{\bf{T}}^{(i)}},{{\bf{s}}^{(i)}}} \right).
\end{aligned}
\end{equation}
Similarly, in the step 4 of $\textbf{Algorithm 1}$, we can obtain the task input bits and time allocation when ${\bf{A}}^{(i+1)}$, ${\bf{B}}^{(i+1)}$, ${\bf{P}}^{(i+1)}$ and ${\bf{s}}^{(i)}$ are given. Herein, we also have 
\begin{equation}
\begin{aligned}
{\cal E}\left( {{{\bf{A}}^{(i + 1)}},{{\bf{B}}^{(i + 1)}},{{\bf{P}}^{(i + 1)}},{{\bf{D}}^{(i)}},{{\bf{T}}^{(i)}},{{\bf{s}}^{(i)}}} \right) \ge \\{\cal E}\left( {{{\bf{A}}^{(i + 1)}},{{\bf{B}}^{(i + 1)}},{{\bf{P}}^{(i + 1)}},{{\bf{D}}^{(i + 1)}},{{\bf{T}}^{(i + 1)}},{{\bf{s}}^{(i)}}} \right).
\end{aligned}
\end{equation}
In the step 5 of $\textbf{Algorithm 1}$, RMS transmissive coefficient can be obtained when ${\bf{A}}^{(i+1)}$, ${\bf{B}}^{(i+1)}$, ${\bf{P}}^{(i+1)}$, ${\bf{D}}^{(i+1)}$ , ${\bf{T}}^{(i+1)}$ are given. Therefore, we have
\begin{equation}
\begin{aligned}
{\cal E}\left( {{{\bf{A}}^{(i + 1)}},{{\bf{B}}^{(i + 1)}},{{\bf{P}}^{(i + 1)}},{{\bf{D}}^{(i + 1)}},{{\bf{T}}^{(i + 1)}},{{\bf{s}}^{(i)}}} \right) \ge\\ {\cal E}\left( {{{\bf{A}}^{(i + 1)}},{{\bf{B}}^{(i + 1)}},{{\bf{P}}^{(i + 1)}},{{\bf{D}}^{(i + 1)}},{{\bf{T}}^{(i + 1)}},{{\bf{s}}^{(i + 1)}}} \right).
\end{aligned}
\end{equation}
Based on the above, we can obtain
\begin{equation}
\begin{aligned}
{\cal E}\left( {{{\bf{A}}^{(i)}},{{\bf{B}}^{(i)}},{{\bf{P}}^{(i)}},{{\bf{D}}^{(i)}},{{\bf{T}}^{(i)}},{{\bf{s}}^{(i)}}} \right) \ge \\{\cal E}\left( {{{\bf{A}}^{(i + 1)}},{{\bf{B}}^{(i + 1)}},{{\bf{P}}^{(i + 1)}},{{\bf{D}}^{(i + 1)}},{{\bf{T}}^{(i + 1)}},{{\bf{s}}^{(i + 1)}}} \right).
\end{aligned}
\end{equation}
which shows that the value of the objective function is non-increasing after each iteration of $\textbf{Algorithm 1}$. Since the objective function must be lower bounded by a finite value, the convergence of $\textbf{Algorithm 1}$ can be guaranteed.

\section{Numerical Results}

In this section, numerical results are provided to evaluate the effectiveness of the proposed joint optimization algorithm in multi-tier computing networks. We consider a three-dimensional (3D) coordinate system in this paper, where the RMS and the DF are located at (0m, 0m, 10m) and (25m, 25m, 10m), respectively, and $K = 5$ users are randomly and uniformly distributed in a circle whose origin is (0m, 0m, 0m) and a radius of 50m. Moreover, DF is equipped with single antenna, and RMS is equipped with $M = 25$ elements. The number of subcarriers is set to 20, i.e., $N = 20$. Assuming that the parameters of all users are the same, we set $W = 1{\rm{ MHz}}$, ${\sigma ^2} =  - 70{\rm{dBm}}$, ${c_t} = {c_r} = {c_m} = {10^3}$ cycles/bit, ${\alpha _t} = {10^{ - 27}}$, ${\alpha _r} = 0.3 \times {10^{ - 27}}$, $P_{\max }^t = P_{\max }^r = 40{\rm{dBm}}$, ${f_c} = 3{\rm{ GHz}}$, ${f_{t,\max }} = 2{\rm{ GHz}}$, ${f_{r,\max }} = 3{\rm{ GHz}}$, and ${f_{m,\max }} = 5{\rm{ GHz}}$ in our numerical simulations \cite{8488502}. Meanwhile, the path loss exponents is set as $\nu=\alpha  = 3$. The path loss with a reference distance of 1m is set to ${C_0} =  - 30{\rm{dB}}$, and we set Rician factor to ${\kappa_1}={\kappa_2} =3{\rm{dB}}$. In addition, the convergence threshold of the proposed algorithm is set to ${10^{ - 3}}$.

We first evaluate the convergence of the proposed joint optimization algorithm. Fig. 3 illustrates the variation of total energy consumption with the number of iterations under different element numbers $M$. One can observe that the total energy consumption for all curves monotonically decreases as the number of iterations increases which means the proposed algorithm can quickly achieve convergence and has good convergence performance. In addition, as the number of RMS elements increases, lower energy consumes, which provides a scheme to reduce system energy consumption by increasing the number of transmissive RMS elements. Hence, the effectiveness and advantages of the transmissive RMS transceiver system are confirmed.

\begin{figure}

	\centerline{\includegraphics[width=9.5cm]{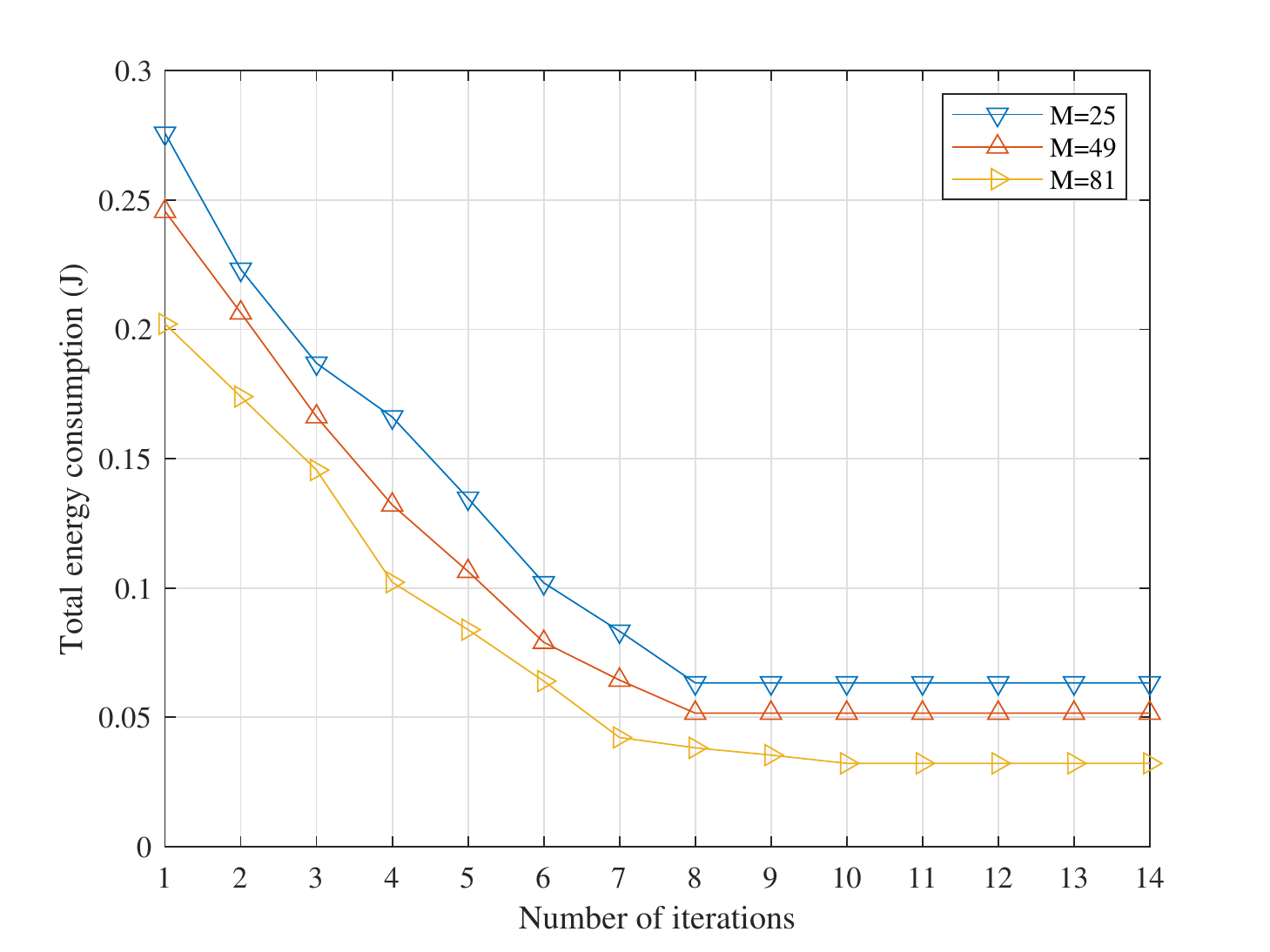}}

	\caption{Convergence behaviour of the proposed joint optimization algorithm.}

	\label{Fig3}

\end{figure}

Then, We compare the performance of our proposed multi-tier computing model with other computing models as following: (1) $\textbf{benchmark 1}$ (i.e., local computing): In this case, users execute all input task bits locally within the time duration $T$. (2) $\textbf{benchmark 2}$ (i.e., partial offloading of computation collaboration): In this case, the task input bits are divided into two parts to be executed at the local and DF relays respectively. (3) $\textbf{benchmark 3}$ (i.e., partial offloading of communication collaboration): In this case, tasks bits are assigned to users and RMS multi-antenna systems to execute, where DF relay assists in offloading. (4) $\textbf{benchmark 4}$ (i.e., RMS-random-phase): In this case, based on the proposed algorithm, the phase of the RMS is not optimized, but a random phase is applied.

We discuss the relationship between the computing capability of different benchmarks and the time duration $T$.  Fig. 4 shows the number of input task bits versus the time duration $T$ under different benchmarks. It can be observed that the number of task input bits under different benchmarks increases as the time duration $T$ increases. That is because the longer time, the more tasks bits can be executed by the CPU. Specifically, the benchmark 2 and benchmark 3 can achieve higher computational capability than the benchmark 1. Furthermore, the benchmark 3 is superior to the benchmark 2 due to the high computation capability of the MEC server equipped with RMS transceiver. In plain sight, the proposed algorithm outperforms other benchmarks and achieves the highest computational capability by leveraging the strengths of both computation collaboration and communication collaboration. In addition, the proposed algorithm performs much better than the RMS-random phase benchmark due to the advantage of the optimized RMS transmissive coefficients.

\begin{figure}
	\centerline{\includegraphics[width=9.5cm]{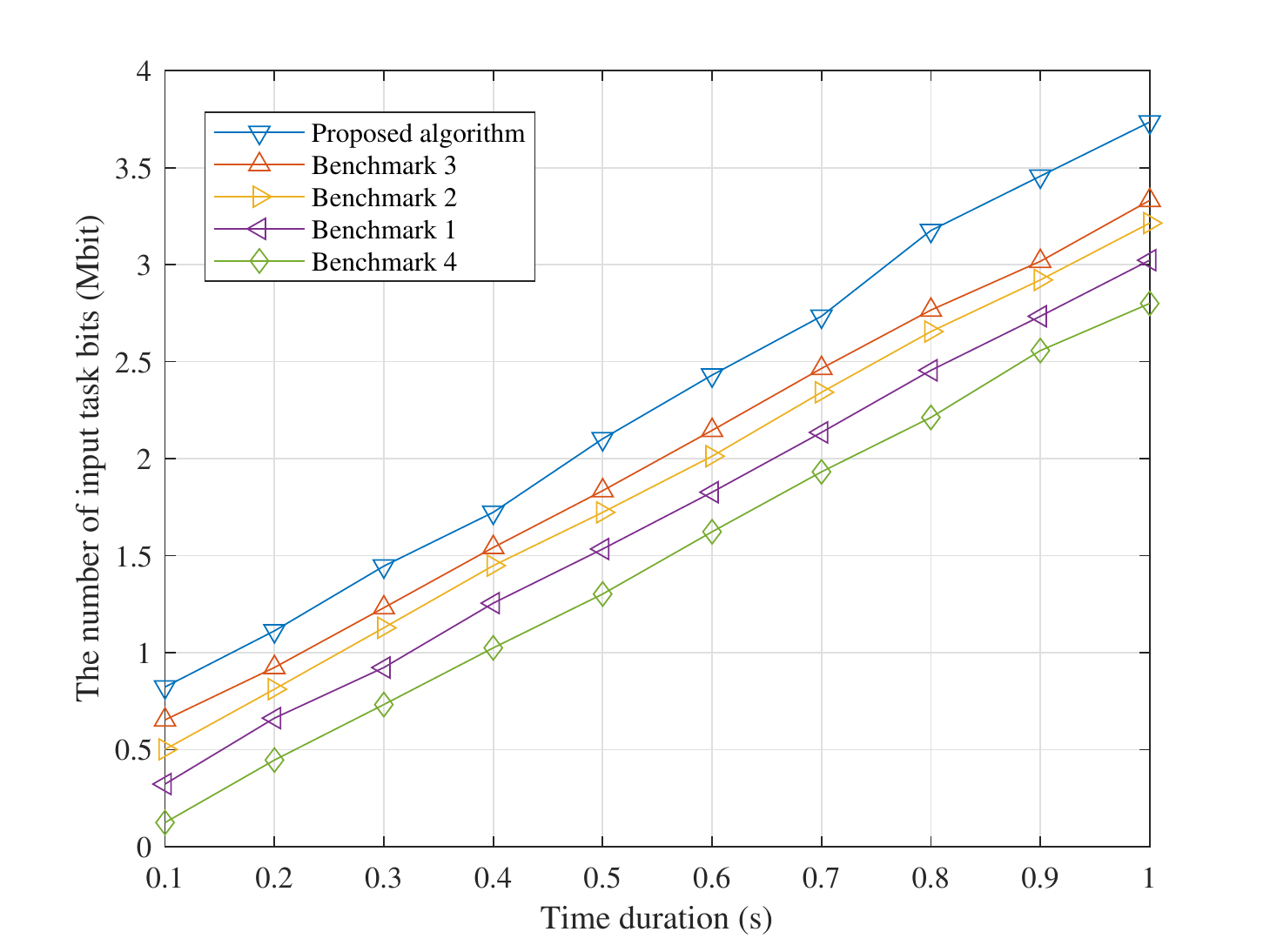}}
	\caption{Number of task input bits versus the time duration $T$ for the proposed algorithm and different benchmark algorithms.}
	\label{Fig4}
\end{figure}

Fig. 5 depicts the variation of the total energy consumption with the length of time duration $T$ for the proposed algorithm and other benchmark algorithms. The total energy consumption for all the benchmraks decreases as time duration $T$ increases. Specific observations are as follows. It can be seen that when the length of time duration $T$ increases, the performance of the proposed algorithm improves and achieves the lowest total energy consumption. Since the transmit power is constant, the offloading time does not vary with $T$ which results in the offloading energy consumption $E_k^{off}$ remaining constant. However, as time increases, more task input bits can be offloaded to the MEC server, and the local computation energy consumption $E_k^{t,comp}$ given in Eq. (29) and computation energy consumption $E_k^{r,comp}$ given in Eq. (32) is subsequently reduced, resulting in a decreasing trend in total energy consumption. This demonstrates the benefits of MEC server for decreasing energy consumption. In addition, the benchmrak 4 performs worst due to the randomness of its phase. For the benchmark 1, as the time duration $T$ increases, according to the expression given in Eq. (29), the local computation energy computation decreases. Meanwhile, both benchmark 2 and benchmark 3 outperform the local computation due to the exploit of computation resources at the DF relay and the transmissive RMS multi-antenna system. 

\begin{figure}

	\centerline{\includegraphics[width=9.5cm]{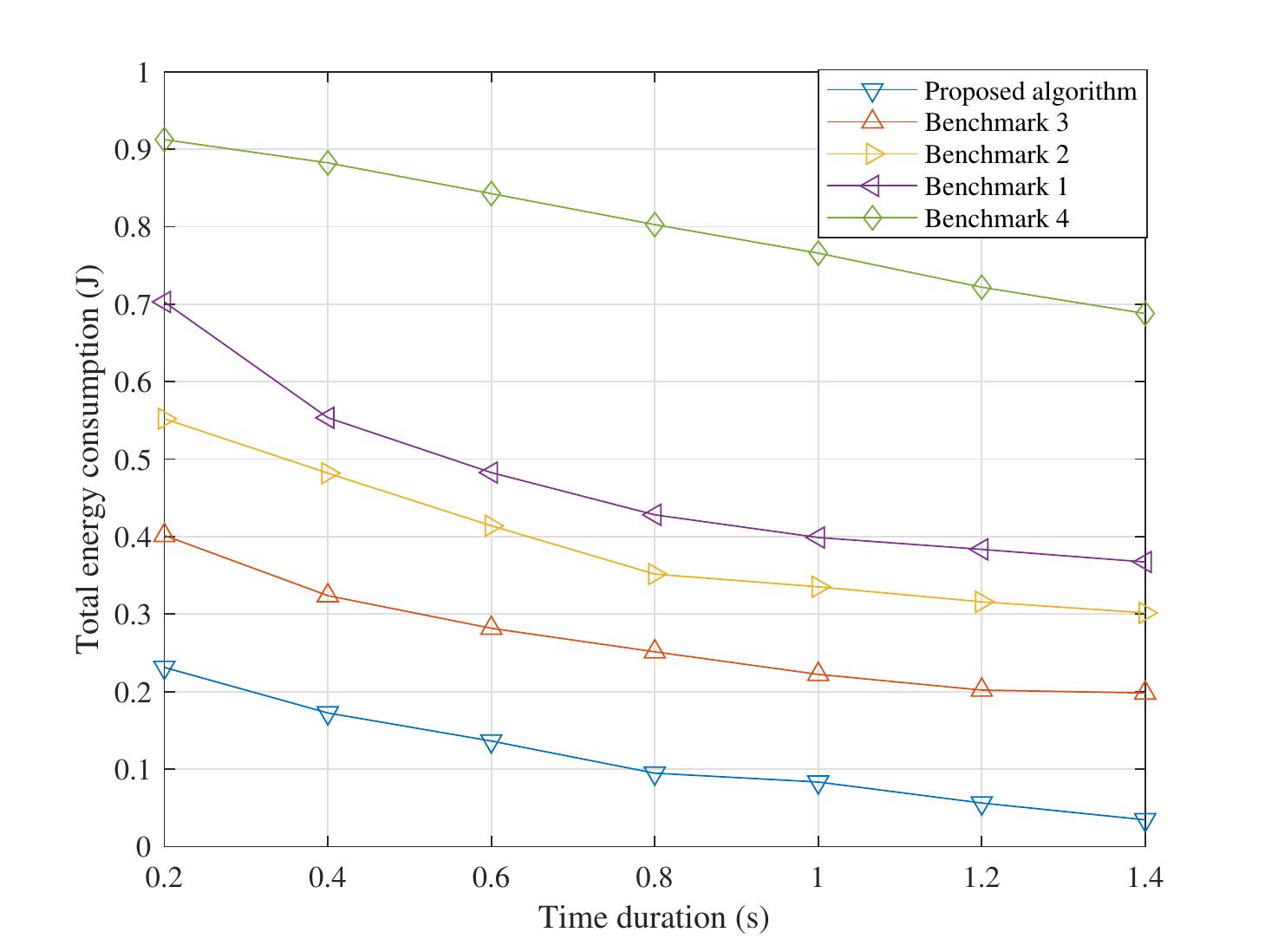}}

	\caption{Total energy consumption versus the time duration $T$ for the proposed algorithm and different benchmark algorithms.}

	\label{Fig5}

\end{figure}

Then, the variation of the total energy consumption with the task input bits $D_k$ is shown in Fig. 6. The results show that the total energy consumption of the proposed algorithm and all benchmarks increases as the number of task input bits increases.  This is because when the number of task input bits increases, the number of local and CN CPU operations increases, resulting in an increase in energy consumption. Furthermore, the proposed algorithm achieves minimum total energy consumption, especially for larger input bits, once again confirming the advantages of the proposed protocol in the multi-tier computing networks in terms of reducing total energy consumption. 

\begin{figure}
	\centerline{\includegraphics[width=9.5cm]{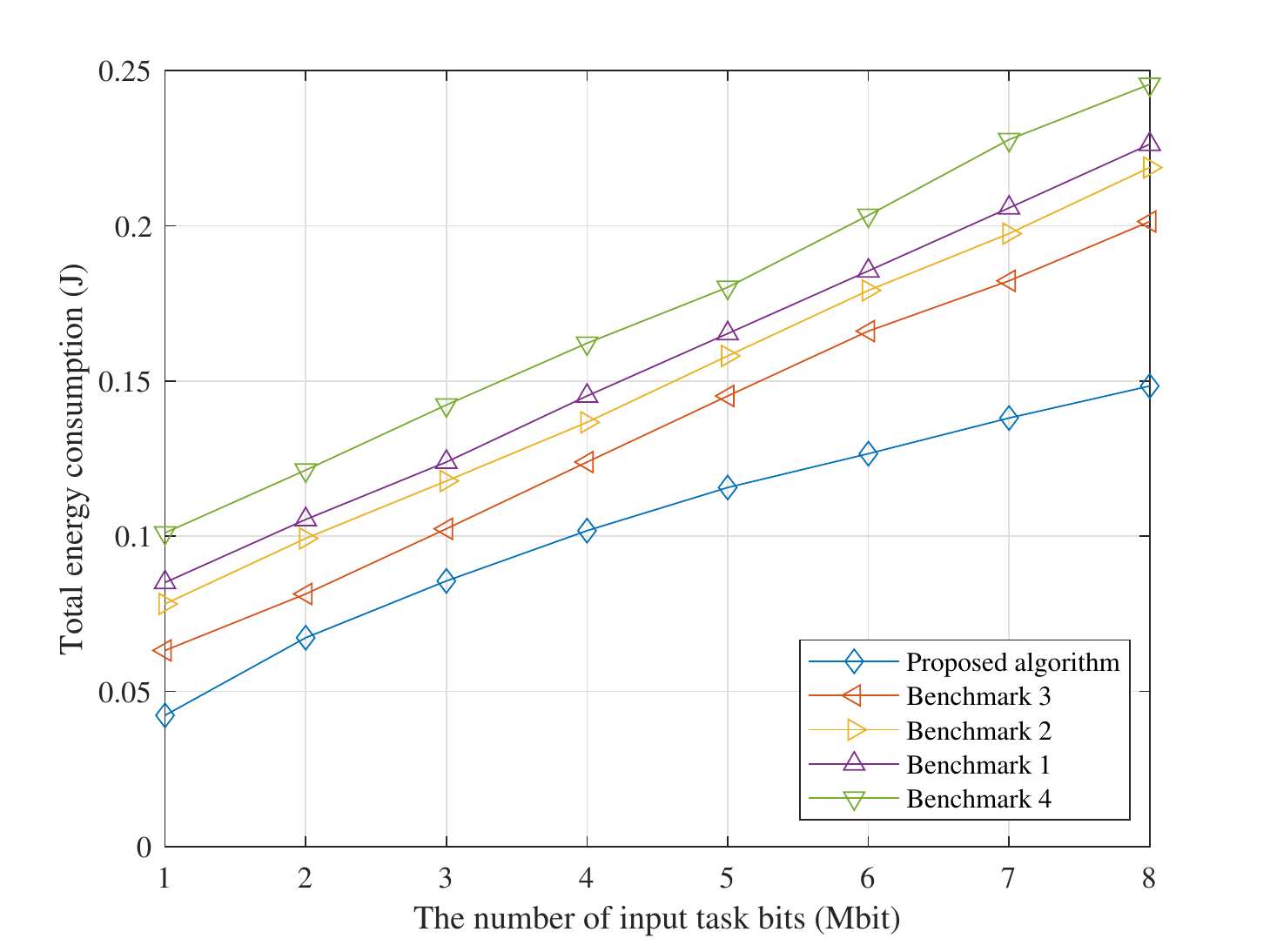}}
	\caption{Total energy consumption versus the task input bits for the proposed algorithm and different benchmark algorithms.}
	\label{Fig6}
\end{figure}







Next, Fig. 7 illustrates the variation of total energy consumption with the number of RMS transmissive elements. Obviously, it is not equipped with RMS transceiver in the benchmark 1 and benchmark 2, so the total energy consumption remains unchanged as the number of the RMS elements increases. We can observe that as the number of RMS elements increases, the total energy consumption decreases for the proposed algorithm and the benchmark 3 and benchmark 4, because the additional elements provide more channel diversity gain, which reduces the total energy consumption by reducing the transmit power.  Specifically, when the number of RMS transmissive elements is the same, the performance of the proposed algorithm is better than the RMS-random-phase benchmark, for the reason that the optimized phase is controllable for the system and the effect of the random phase on the system is uncontrollable. In addition, the benchmark 3 performs less well than the proposed algorithm since lack of computing resources for DF relay. Meanwhile, the proposed algorithm achieves the lowest total energy consumption among all the benchmarks. Therefore, increasing the number of RMS transmissive elements is of great value for practical use.

\begin{figure}

	\centerline{\includegraphics[width=9.5cm]{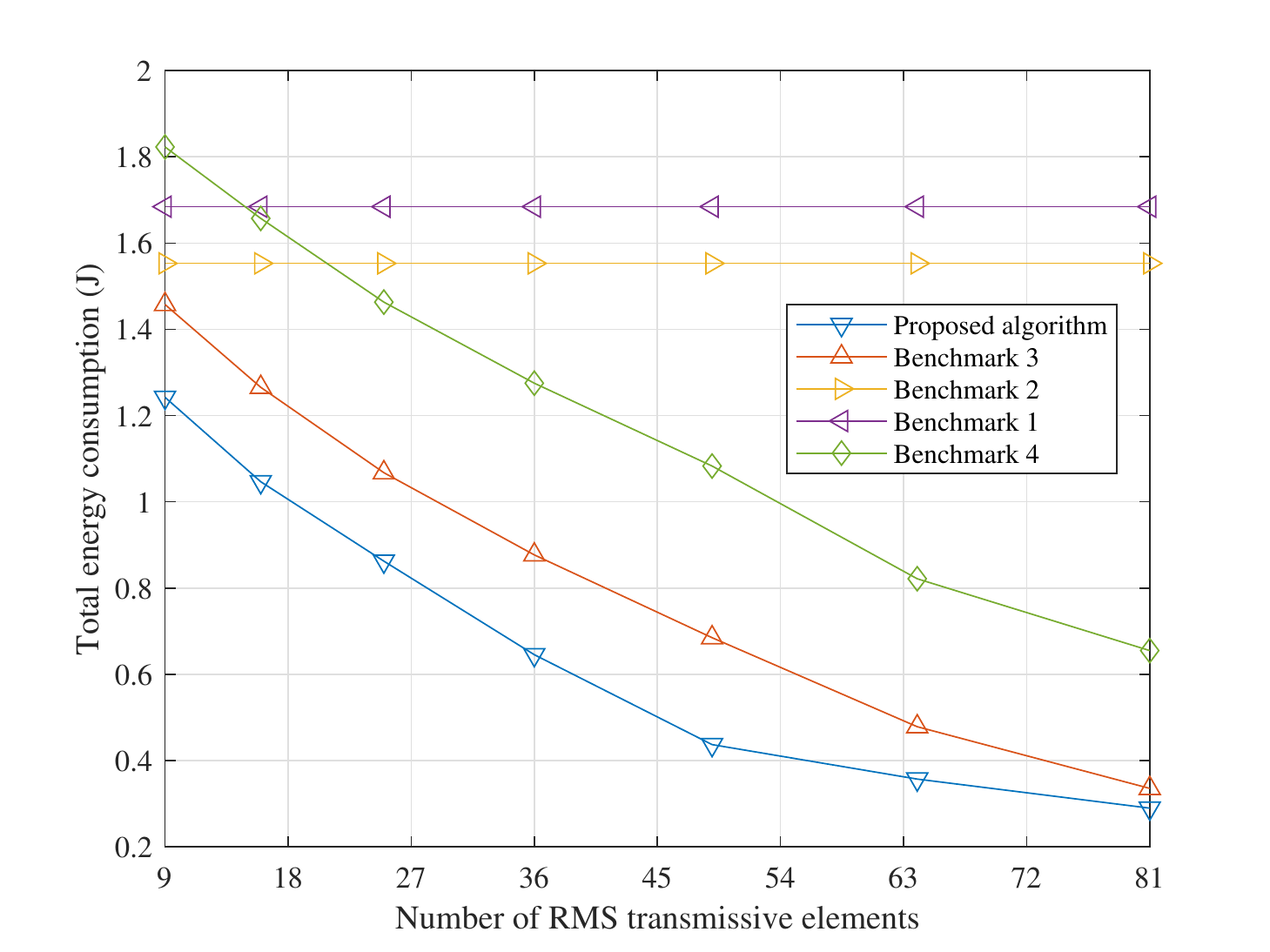}}

	\caption{Total energy consumption versus the number of RMS transmissive elements for the proposed algorithm and different benchmark algorithms.}

	\label{Fig7}

\end{figure}






Fig. 8 elaborates the variation of total energy consumption with the number of users. It can be seen from Fig. 9 that when the number of users increases, the total energy consumption increases. The reason is that as the number of users increases, the total input task bits increases, resulting in increased energy consumption for  offloading and computation. When the number of users is the same, the proposed algorithm outperforms other benchmarks. The specific reasons are similar to those mentioned above, and are omitted here.
\begin{figure}
	\centerline{\includegraphics[width=9.5cm]{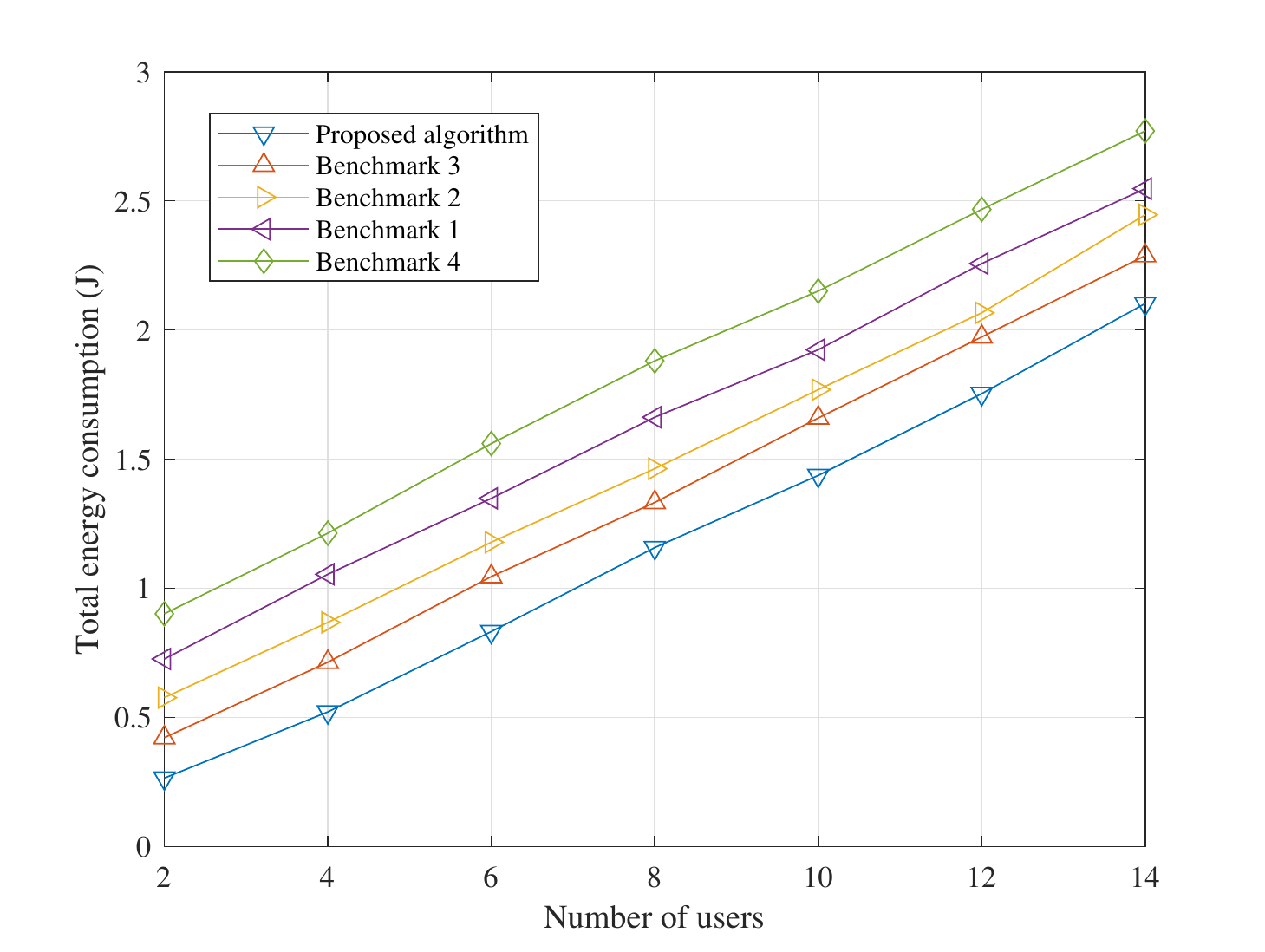}}
	\caption{Total energy consumption versus the number of users for the proposed algorithm and different benchmark algorithms.}
	\label{Fig8}
\end{figure}

Afterwards, we evaluate the performance of our proposed joint optimization algorithm compared with other benchmark algorithms as follows. (4) $\textbf{benchmark 4}$ (i.e., RMS-random-phase): In this case, based on the proposed algorithm, the phase of the RMS is not optimized, but a random phase is applied. (5) $\textbf{benchmark 5}$ (i.e., RMS-SDR-phase): In this case, based on the proposed algorithm, the phase of the RMS is is optimized by implementing semidefinite relaxation (SDR) technique. (6) $\textbf{benchmark 6}$ (i.e., three-stage algorithm): In this case, the three subproblems are optimized based on the proposed algorithm, but no alternating optimization is performed. (7) $\textbf{benchmark 7}$ (i.e., upper bound): In this case, after solving sub-problem 1 to obtain the subcarrier assignment variable, it is not restored to a discrete binary variable.

Fig. 9 shows the variation of task input bits with time duration $T$ under different optimization algorithms. It can be seen that, as the time duration $T$ increases, since the CPU can execute more offloading tasks, the number of task input bits increases. In addition, within the same time duration $T$, the performance of benchmark 7 is better than our proposed algorithm and other benchmark algorithms. This is because after solving sub-problem 1, the benchmark algorithm does not approximately restore the sub-carrier allocation variables, which ensures that the algorithm performance. In addition, the performance of the proposed algorithm is better than benchmark 5 because the adopted DC algorithm has better performance guarantee than the SDR algorithm. Compared with the proposed algorithm, benchmark 6 does not perform alternating optimization after solving the three sub-problems, so it is easy to fall into local optimum and cannot achieve global optimum, so its performance is poor. 
\begin{figure}
	\centerline{\includegraphics[width=9.5cm]{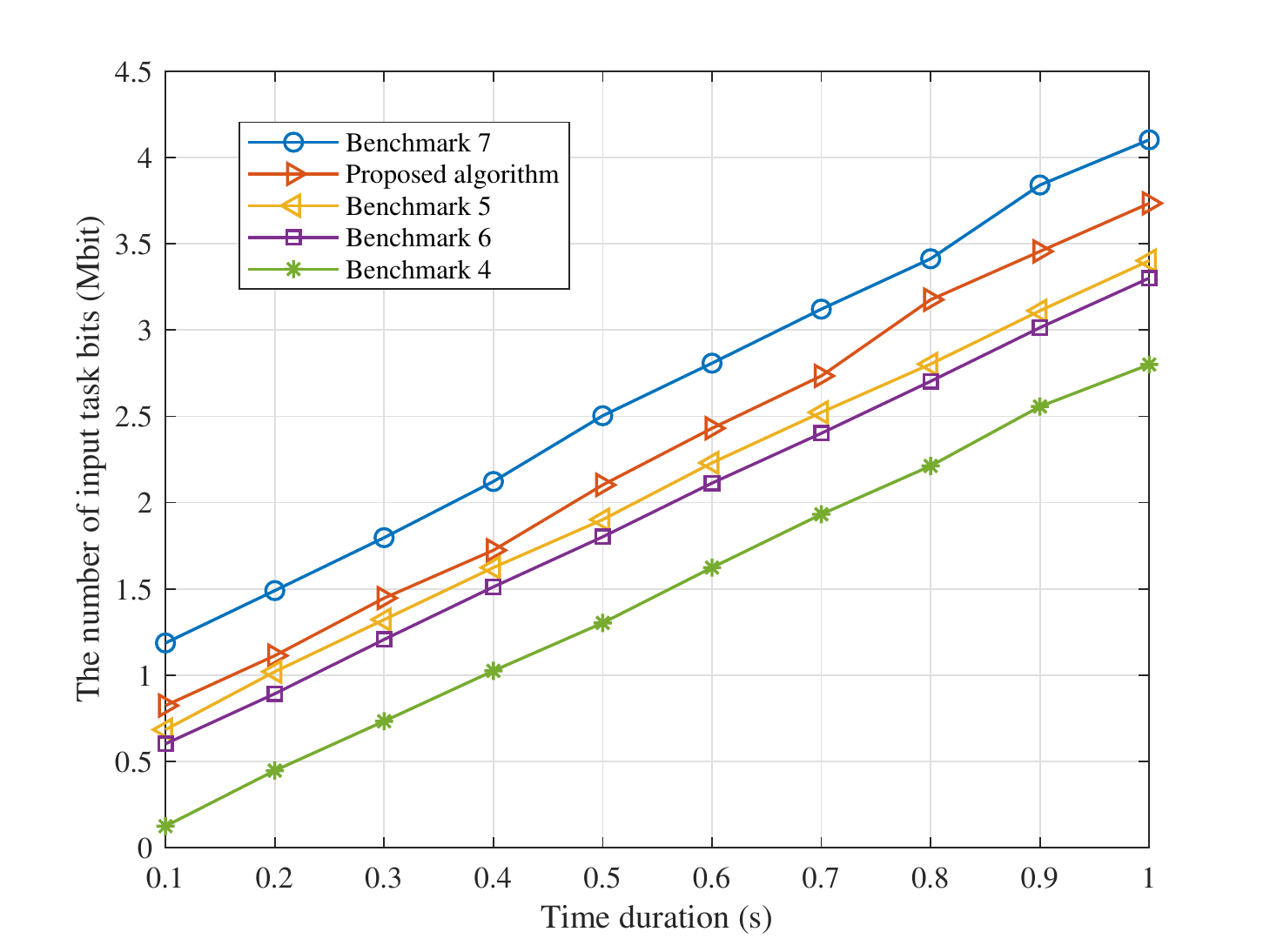}}
	\caption{Number of task input bits versus the time duration $T$ for the proposed algorithm and different benchmark algorithms.}
	\label{Fig9}
\end{figure}

Finally, Fig. 10 illustrates the variation of total energy consumption with the number of task input bits under different optimization algorithms. It can be observed that when the number of task input bits increases, the total energy consumption also increases. When the number of task input bits is the same, the performance of benchmark 7 is better than that of the proposed algorithm and other benchmark algorithms, and the performance of the proposed algorithm is also better than that of other benchmark algorithms. The specific reasons are similar to the above, and are omitted here.
\begin{figure}
	\centerline{\includegraphics[width=9.5cm]{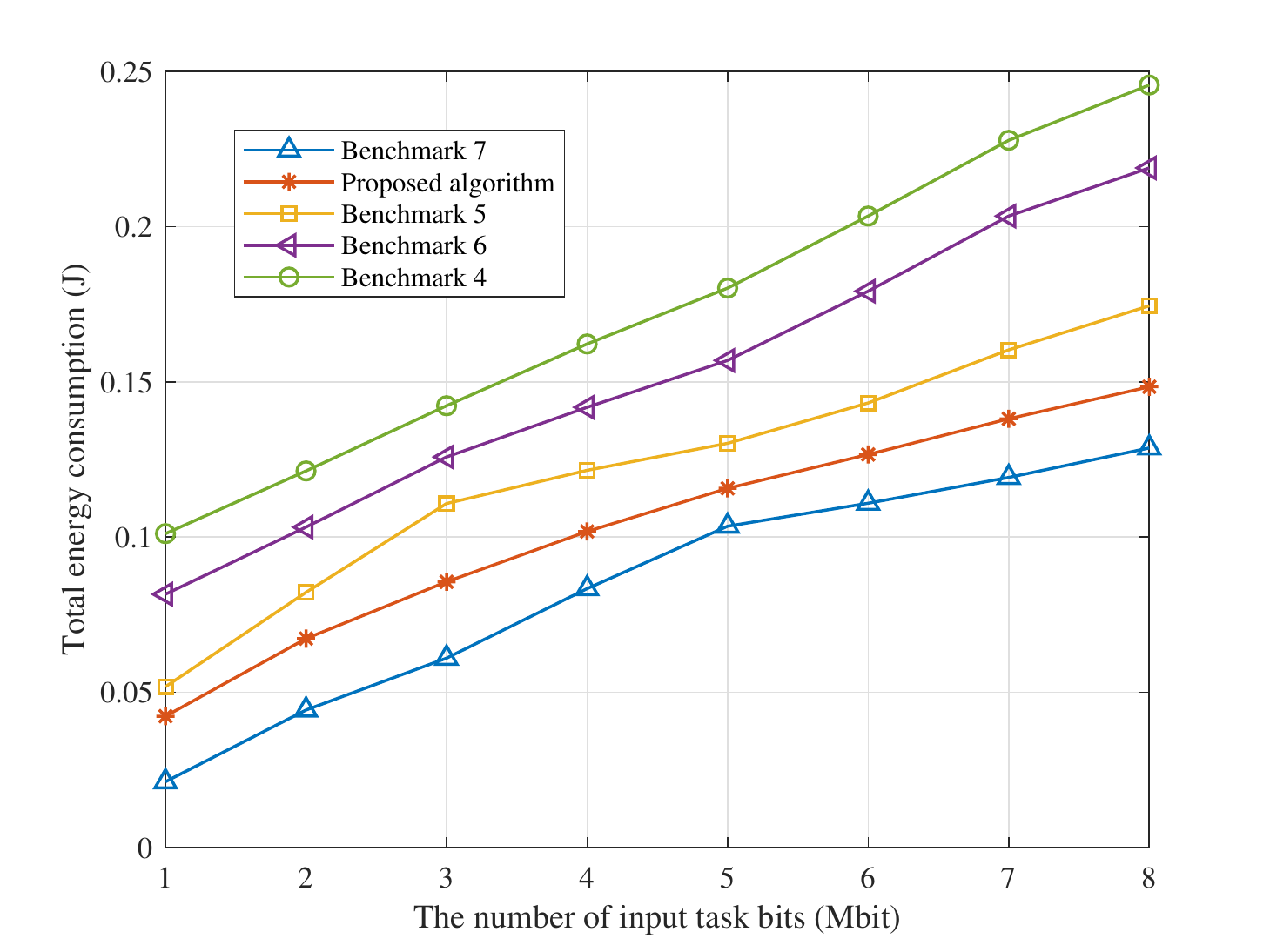}}
	\caption{Total energy consumption versus the task input bits for the proposed algorithm and different benchmark algorithms.}
	\label{Fig10}
\end{figure}

\section{Conclusions}
This paper investigates the total energy consumption minimization problem of transmissive RMS transceiver enabled multi-tier computing networks. Specifically, under the constraints of the computing and energy resources, subcarrier allocation, input task bits allocation, time slot allocation, user and DF relay transmit power allocation, and the RMS transmissive coefficients are jointly optimized. First, we transform the problem into a tractable problem. Then, in order to solve the transformed problem, we apply the BCD algorithm framework to divide the original problem into three sub-problems for solving. Given the other variables, we solve the variables to be optimized through SCA, DC programming, etc., and then alternately optimize the three sub-problems until convergence is achieved. Then, the computational complexity and convergence analysis of the proposed algorithm are given. Finally, the numerical simulation results verify the convergence and effectiveness of the proposed algorithm, which illustrates that the proposed algorithm is capable of decreasing the total energy consumption to a large extent, and the advantages of utilizing transmissive RMS in this architecture for energy reduction are obvious.

\ifCLASSOPTIONcaptionsoff
  \newpage
\fi


\bibliographystyle{IEEEtran}
\bibliography{reference}
\end{document}